\RequirePackage[framemethod=default]{mdframed}
\RequirePackage{tcolorbox}
\RequirePackage[parfill]{parskip}
\RequirePackage{color,soul}
\RequirePackage[svgnames]{xcolor}
\RequirePackage{colortbl}

\documentclass[acmsmall]{acmart}

\usepackage{comment}
\usepackage{enumitem}
\usepackage{xurl}
\AtBeginDocument{%
  }

\setcopyright{acmlicensed}
\copyrightyear{2018}
\acmYear{2018}
\acmDOI{XXXXXXX.XXXXXXX}

\acmJournal{JACM}
\acmVolume{37}
\acmNumber{4}
\acmArticle{111}
\acmMonth{8}

\begin{document}

\title{A Comprehensive Study of Bugs in Modern Distributed Deep Learning Systems}


\author{Xiaoxue Ma}
\email{kxma@hkmu.edu.hk}
\orcid{0000-0002-5476-6074}
\affiliation{%
  \institution{Hong Kong Metropolitan University}
  \city{Hong Kong}
  \country{China}
}

\author{Wanwei Zhan}
\email{zww1023004495@whut.edu.cn}
\affiliation{%
  \institution{Wuhan University of Technology}
  \city{Wuhan}
  \country{China}
}

\author{Jiale Chen}
\email{355224@whut.edu.cn}
\affiliation{%
  \institution{Wuhan University of Technology}
  \city{Wuhan}
  \country{China}
}
\author{Yishu Li}
\email{sliy@hkmu.edu.hk}
\orcid{xx}
\affiliation{%
  \institution{Hong Kong Metropolitan University}
  \city{Hong Kong}
  \country{China}
}
\author{Jacky Keung}
\email{jacky.keung@cityu.edu.hk}
\orcid{0000-0002-3803-9600}
\affiliation{%
  \institution{City University of Hong Kong}
  \city{Hong Kong}
  \country{China}
}
\author{Federica Sarro}
\email{f.sarro@ucl.ac.uk}
\orcid{0000-0002-9146-442X}
\affiliation{%
  \institution{University College London}
  \city{London}
  \country{UK}
}

\renewcommand{\shortauthors}{Trovato et al.}

\begin{abstract}
In today’s data-driven era, deep learning is vital for processing massive datasets, yet single-device training is constrained by computational and memory limits. Distributed deep learning overcomes these challenges by leveraging multiple GPUs or machines in parallel. While general-purpose frameworks (e.g., TensorFlow and PyTorch) provide distributed capabilities, these are often add-on features that demand significant manual effort for advanced parallelism, underscoring the need for specialized frameworks.
This study conducts the first large-scale empirical analysis of practitioner challenges in dedicated distributed frameworks. We examine 849 real-world issues from DeepSpeed, Megatron-LM, and Colossal-AI and construct a taxonomy of 34 bug symptoms, 28 root causes, and 6 fix patterns. Crucially, we establish explicit mappings between symptoms, causes, and fixes across distributed training stages, enabling a systematic understanding of how issues emerge and are resolved. Our results show that 45.1\% of bug symptoms are unique to distributed frameworks, with setup failures, memory issues, and performance anomalies being the most prevalent. Moreover, 95\% of issues in the communication setup stage occur exclusively in distributed contexts. We also find over 60\% of cases can be resolved through version and dependency management, and distributed feature, API, and communication tuning. Based on these findings, we provide actionable implications. 
\end{abstract}




\maketitle

\section{Introduction}

Deep learning \cite{lecun2015deep,yang2022survey} is a foundational technology in software engineering, driven by the ever-growing volume of data generated across industries. Large language models are particularly prominent due to their capacity for understanding and extracting insights from massive datasets. However, these models often contain billions of parameters, presenting formidable computational and memory demands that far exceed the capacity of any single device.
Non-distributed deep learning confines all data, parameters, and computations to a single device (e.g., one GPU or CPU), thereby restricting the size and speed of model training. To surpass these limitations, distributed deep learning employs multiple GPUs or machines, distributing data and computations and enabling large-scale parallelism \cite{borzunov2023distributed,shanahan2024talking}. This not only accelerates training but also makes possible the development and deployment of sophisticated models that would otherwise be infeasible.

Existing studies such as Liu et al.~\cite{liu2023rise} have analyzed practitioner challenges in distributed training with general-purpose frameworks like TensorFlow \cite{abadi2016tensorflow}, PyTorch \cite{paszke2019pytorch}, and Keras \cite{gulli2017deep}. These frameworks are widely adopted for their flexibility and versatility, providing a comprehensive suite of tools for model building, training, and inference across diverse hardware platforms from CPUs to multi-GPU clusters. Although they support distributed training, this capability forms just one part of their broader design, which primarily emphasizes extensibility, ease of use, and coverage of a broad range of deep learning tasks. Consequently, users benefit from a rich ecosystem and user-friendly features, but distributed functionality is not as deeply integrated or optimized as in specialized frameworks \cite{gao2020estimating,dai2022reveal}. 
In contrast, 
purpose-built distributed deep learning frameworks such as DeepSpeed~\cite{rasley2020deepspeed}, Megatron-LM~\cite{shoeybi2019megatron}, and Colossal-AI~\cite{li2023colossal})
are explicitly optimized for training models at scale, integrating advanced parallelization technologies (e.g., data, model, and pipeline parallelism) and ZeRO (Zero Redundancy Optimizer)\cite{rajbhandari2020zero}, which minimizes memory redundancy by partitioning optimizer states and parameters across devices. Unlike general-purpose frameworks, these distributed frameworks provide highly efficient, integrated solutions, streamlining the training of massive models that would be impractical to implement using manual configuration or standard distributed capabilities. However, while these systems deliver unparalleled scalability and efficiency, they also introduce new and distinctive reliability and integration challenges.

In this work we provide the first comprehensive empirical study of practitioner issues in dedicated distributed deep learning frameworks.  
To this end, we construct the first large-scale empirical dataset of 849 real-world practitioner issues from GitHub, providing unique insights into modern distributed training and inference. Understanding these user-reported challenges will enable framework providers to develop targeted solutions that help practitioners prevent, detect, and resolve problems more effectively. We structure our study into the following research questions (RQs):

\textbf{RQ1: Bug Symptoms:} 
What are the symptoms of bugs frequently encountered in specialized distributed frameworks (DeepSpeed, Megatron-LM, Colossal-AI)? 
To answer RQ1, we devise a comprehensive taxonomy of bug symptoms throughout distributed deep learning, finding that 45.1\% are unique to distributed frameworks. We found that the most prevalent distributed-specific symptoms are initialization and launcher setup failures, memory issues, and unexpected performance.

\textbf{RQ2: Root Causes of Bugs:}
What are the underlying root causes of bugs in distributed deep learning?
To answer RQ2, we systematically examine each reported issue and its discussion, identifying and categorizing the root causes at every stage of distributed deep learning and mapping these causes to their respective symptoms via heatmaps. We found that the most frequent root causes are configuration errors, API misuse, mismatched or incompatible components, and incorrect implementation.
Notably, during communication setup, 95\% of issues are unique to distributed frameworks, such as node rank mismatches, NCCL or multi-GPU failures, protocol-related hangs, and resource allocation errors. Based on these findings, we recommend actionable practices for framework providers and users, such as using simple SSH, checking network and dependencies, and verifying memory settings to increase reliability and efficiency.

\textbf{RQ3 Fix Patterns of Bugs:} What are the common patterns to fix bugs in distributed deep learning? 
To answer RQ3, we examine fix patterns for each issue, correlating them with the identified root causes from RQ2. Over 90\% of distributed deep learning issues can be resolved with just four main fix strategies. We found that the most significant fixes are managing versions and dependencies (37\% of cases) and tuning distributed APIs, communication, and features (32\%). We further quantify bug resolution effort by examining issue open duration and the number of comments, providing examples to illustrate resolution complexity and strategies. For further improvement, we recommend that framework providers deliver clear guidance, sample configurations, and routinely check compatibility and troubleshooting resources to help users address frequent challenges (e.g., integration) efficiently. 

This study presents a comprehensive set of findings and practical insights for both framework developers and practitioners in distributed deep learning. Table~\ref{table:1} summarizes the key findings and implications. The main contributions of this study are as follows.

\begin{itemize}[leftmargin=12pt, labelsep=5pt] 
    \item A first large-scale empirical analysis of issues in distributed deep learning systems, rigorously examining 849 bugs across three major distributed deep learning frameworks. Our annotated dataset is made publicly available to support future research \cite{Bugissue}.
    \item An analysis of symptoms and causes of bugs in each stage of distributed deep learning, especially those unique to distributed frameworks. We also identify low-effort fixing strategies that can be adopted to address these issues.
    \item A quantitative assessment of bug fixing effort and typical fix strategies using representative real-world examples, offering valuable insights into the practical challenges faced by developers working with distributed deep learning frameworks.
\end{itemize}

\begin{figure}[!t]
   \centering
   \includegraphics[width=\linewidth]{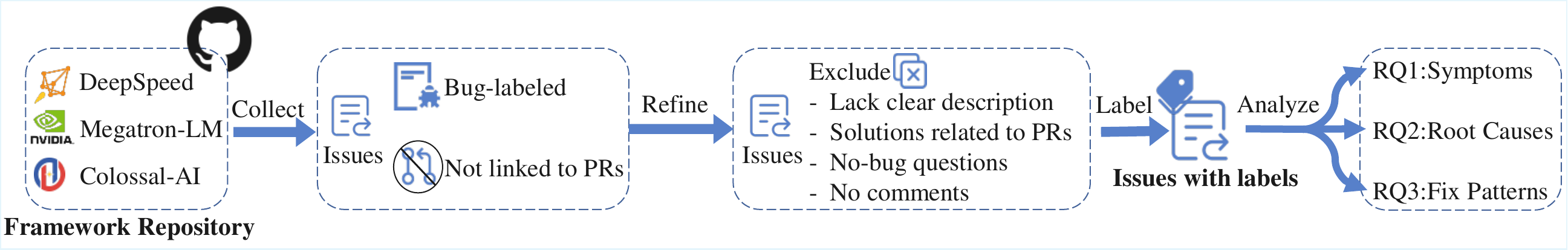}
   \vspace{-0.8cm}
   \caption{Workflow of our study.}
   \label{fig:framework}
   \vspace{-0.6cm}
\end{figure}

\section{Methodology}
To characterize developers' challenges in distributed training and inference, we collected and analyzed GitHub issues, with the methodology visualized in Figure~\ref{fig:framework}.

\subsection{Data Collection}
General deep learning frameworks such as TensorFlow, PyTorch, and Keras provide foundational support for distributed training through modules like PyTorch’s \texttt{torch.distributed} and TensorFlow’s \texttt{tf.distribute}. However, these frameworks are inherently general-purpose, designed to accommodate a broad spectrum of machine learning workflows. 
Their distributed capabilities are typically implemented as extensions to single-node architectures, requiring substantial manual configuration to implement advanced parallelism strategies (e.g., model parallelism). Consequently, while suitable for basic distributed scenarios, they often encounter scalability limitations when applied to large-scale models with billions or trillions of parameters.
In contrast, our study focuses on three specialized frameworks (i.e., DeepSpeed \cite{rasley2020deepspeed}, Megatron-LM \cite{shoeybi2019megatron}, and Colossal-AI \cite{li2023colossal}) engineered explicitly to address the challenges of large-scale parallelism in deep learning. These frameworks incorporate optimized techniques including model parallelism, which divides the model into segments for different devices; data parallelism, where each device maintains an identical copy of the model to process distinct subsets of training data; pipeline parallelism, which segments the model into sequential stages for processing micro-batches across devices; and ZeRO, which minimizes resource duplication by distributing optimizer states, gradients, and model parameters across devices. Together, these techniques enable efficient distribution of computations and memory usage across multiple nodes, facilitating the training of large-scale deep learning models. 
The three frameworks simplify the implementation of complex parallelism strategies, making it possible to train models that would be infeasible with general frameworks. 
Their widespread adoption in high-performance computing and large language model research \cite{aminabadi2022deepspeed, li2024deepspeed, narayanan2021efficient} ensures that issues reported reflect critical challenges specific to distributed deep learning. 
Bugs encountered during the utilization of these frameworks often arise from challenges inherent to distributed deep learning workflows, such as inter-node synchronization failures, NCCL (NVIDIA Collective Communications Library) timeouts, and distributed memory issues (e.g., GPU OOM (out-of-memory) errors associated with ZeRO-3 optimization). 
By systematically analyzing these challenges, our research offers actionable insights aimed at improving the reliability and scalability of distributed deep learning frameworks.

\textbf{Data Source Selection.}
Consistent with prior work \cite{islam2019comprehensive,islam2020repairing,  liu2023rise,shah2025towards}, we mined issues posted on GitHub and Stack Overflow, two widely adopted data sources for studying software-related problems. 
On Stack Overflow, initial tag-based searches \cite{cao2022understanding, liu2023rise, wang2023compatibility, islam2019comprehensive, humbatova2020taxonomy} yielded only 29 DeepSpeed posts and none for the other two frameworks. A keyword-based search expanded these results to 108 (DeepSpeed), 6 (Megatron-LM), and 1 (Colossal-AI). Given the substantial disparity in volume between GitHub and Stack Overflow data, we focused exclusively on GitHub issues.
For each of the selected framework, GitHub hosts corresponding repositories
where practitioners’ issues encountered during framework usage are documented. 
To ensure we focus on community-acknowledged, non-obsolete bugs specific to framework usage, we applied two criteria: (1) selecting only closed ``bug''-labeled issues (following previous work \cite{humbatova2020taxonomy}) to exclude outdated discussions and unverified reports; and (2) excluding issues with linked pull requests to isolate bugs encountered during distributed deep learning rather than internal development.
After Step 1, we identified 946, 123, and 500 issues in DeepSpeed, Megatron-LM, and Colossal-AI, respectively.

\textbf{Dataset Construction.} 
Two authors independently reviewed all issues extracted in Step 1 to refine the final dataset. We excluded issues that (1) lacked a clear and detailed description, making it impossible for practitioners or framework developers to reproduce the bug, (2) had solutions related to internal framework fixes, such as codebase modifications, (3) were non-bug queries (e.g., implementation questions), or (4) received no comments, indicating no community validation. 
For example, DeepSpeed issue \#\texttt{1342} \cite{deepspeed_issue1342} was excluded for ambiguity: the disk OOM error lacked contextual details (e.g., during checkpoint saving or regular training), making reproduction impossible.
Issue \#\texttt{1364} \cite{deepspeed_issue1364}was excluded as it was resolved by linking to a merged Pull Request \#\texttt{1362}\cite{deepspeed_issue1362}, which removed the \texttt{max\_seq\_length} parameter from both the documentation and \texttt{DeepSpeedTransformerConfig} initialization, reflecting an internal codebase adjustment rather than a user-facing bug. The disagreements were resolved by a third author. Inter-rater reliability was measured using Cohen’s Kappa coefficient \cite{viera2005understanding, mchugh2012interrater}, calculated by having two authors independently label the same set of data and assign labels. The resulting coefficient reached 0.86, indicating a strong agreement \cite{yang2025towards}. The final dataset comprises 849 bug-related issues concerning the use of distributed deep learning systems (496 from DeepSpeed, 48 from Megatron-LM, and 305 from Colossal-AI). This scale is notably comprehensive compared to that of similar studies on traditional deep learning systems: 224 performance problems \cite{cao2022understanding}, 352 compatibility issues \cite{wang2023compatibility}, 365 silent bugs \cite{hong2024investigating}, and 569 distributed training issues \cite{liu2023rise}.

\subsection{Manual Labeling}
We manually examined all bug-related issues in our dataset to systematically identify symptoms, root causes, and common fix patterns. Following prior work \cite{liu2023rise}, we reconstructed a hierarchical taxonomy of these bugs, organizing them according to the programming pipeline of distributed deep learning systems, which includes five phases: Package Build and Import, Communication Setup, Data and Model Preparation, Training and Evaluation, and Inference. Our review revealed that existing stage-specific labels (e.g., ``training'' or ``inference'') were often incomplete or inaccurate.
A representative example is DeepSpeed issue \#\texttt{1766}\cite{deepspeed_issue1766}, which was labeled only as ``bug'' despite clearly involving an inference-specific problem. The issue described a CUDA OOM error occurring during inference initialization via \texttt{deepspeed.init\_inference} (an inference-specific API) in a 4-GPU distributed setup. Since no references to training operations were present in the issue, we properly classified it as an inference-related bug.
To establish an overview of the dataset, we first conducted a pilot labeling process using 50\% of the data. This subset was used by the authors to familiarize themselves with distributed deep learning issues, during which initial pilot taxonomies were developed. Subsequently, the remaining 50\% of the dataset was employed to validate the taxonomies through five iterative rounds, with the taxonomies refined continuously based on insights from each validation cycle.

\textbf{Pilot Labeling.}
First, 50\% of the dataset was randomly sampled for pilot labeling. Two authors independently applied an open coding procedure to develop categories for symptoms, root causes, and fix patterns by analyzing the sampled bug-related issues. Specifically, they carefully reviewed all issues to contextualize each case, then assigned a set of labels to describe three dimensions:
(1) \textit{symptoms} refer to observable manifestations of the bug, including its specific appearance and the distributed deep learning stage where it occurs; (2) \textit{root causes} denote the underlying reasons for the bug, whether explicitly stated or reasonably inferred from the issue descriptions; and (3) \textit{fix patterns} encompass methods by which the bug was resolved (e.g., solutions confirmed by the issue submitter) or could be resolved (e.g., plausible solutions proposed by contributors). 
Root causes and fix patterns are optional: issues lacking explicit information for these dimensions were labeled as \textit{Unknown}. 
For example, GitHub Issue \#\texttt{1460}\cite{deepspeed_issue1460} reported an error occurring during matrix multiplication with \texttt{SparseSelfAttention}. Contributor A (a framework developer) attributed the error to recent module structure changes, and Contributor B (a practitioner) independently confirmed the error and proposed a verified solution (updating the import path). 
Despite the original issue submitter not responding, the issue was classified as resolved because the solution was reproducible and aligned with existing fixes. 

Following individual labeling, we adopted a rigorous bottom-up approach \cite{vijayaraghavan2003bug, humbatova2020taxonomy} to develop hierarchical taxonomies. Similar labels were systematically grouped into distinct categories, which were then organized under broader parent categories while strictly maintaining an ``is a'' hierarchical relationship (e.g., ``Incorrect Device Configuration is a Setup Failure''). This taxonomy development occurred through an iterative consensus process: two authors reviewed and refined each version, with any disagreements resolved through discussion with a third author. The process continued until full consensus was achieved on all labels, categories, and taxonomic structures. Finally, all authors conducted an online validation meeting to thoroughly examine the complete taxonomy, including reviewing individual labels, category mappings, and hierarchical relationships, and made necessary minor adjustments to ensure coherence and completeness. 


\textbf{Taxonomy Validation.}
To validate the initial taxonomies, two authors independently labeled the remaining 50\% of bug-related issues, classifying each into symptom, root cause, and fix pattern categories using the same criteria as the pilot labeling phase. Issues that could not be categorized within the existing taxonomies were labeled under the \textit{Undecided} category, accompanied by new labels for the three dimensions.
The validation process consisted of five rounds, with each round analyzing 20\% of the remaining dataset. In each round, inter-rater agreement for the independent labels was assessed. With input from the third author, all authors collaboratively resolved labeling conflicts, reviewed \textit{Undecided} issues with new labels, and determined whether new categories should be integrated into the taxonomies. Following these adjustments, all \textit{Undecided} issues were reclassified into the revised taxonomies. 
In total, \textbf{16, 18, and 12} new categories were added to the symptom, root cause, and fix pattern taxonomies, respectively, with an average Cohen’s Kappa coefficient of 0.84 across all dimensions. 
In particular, new categories were added to the symptom, root cause, and fix pattern taxonomies with an average Cohen’s Kappa coefficient of 0.84 across all dimensions. No new categories were added in the final round, indicating that the taxonomies had reached saturation and could classify all issues comprehensively.
Using these real-world validated issues, we address Research Questions (RQs) 1–3 in Section~\ref{sec:results}.

\section{Results}
\label{sec:results}

\subsection{Symptoms (RQ1)}
\label{sec:sym}

\begin{figure}[!tb]
\vspace{-0.8cm}
   \centering
   \includegraphics[width=\linewidth]{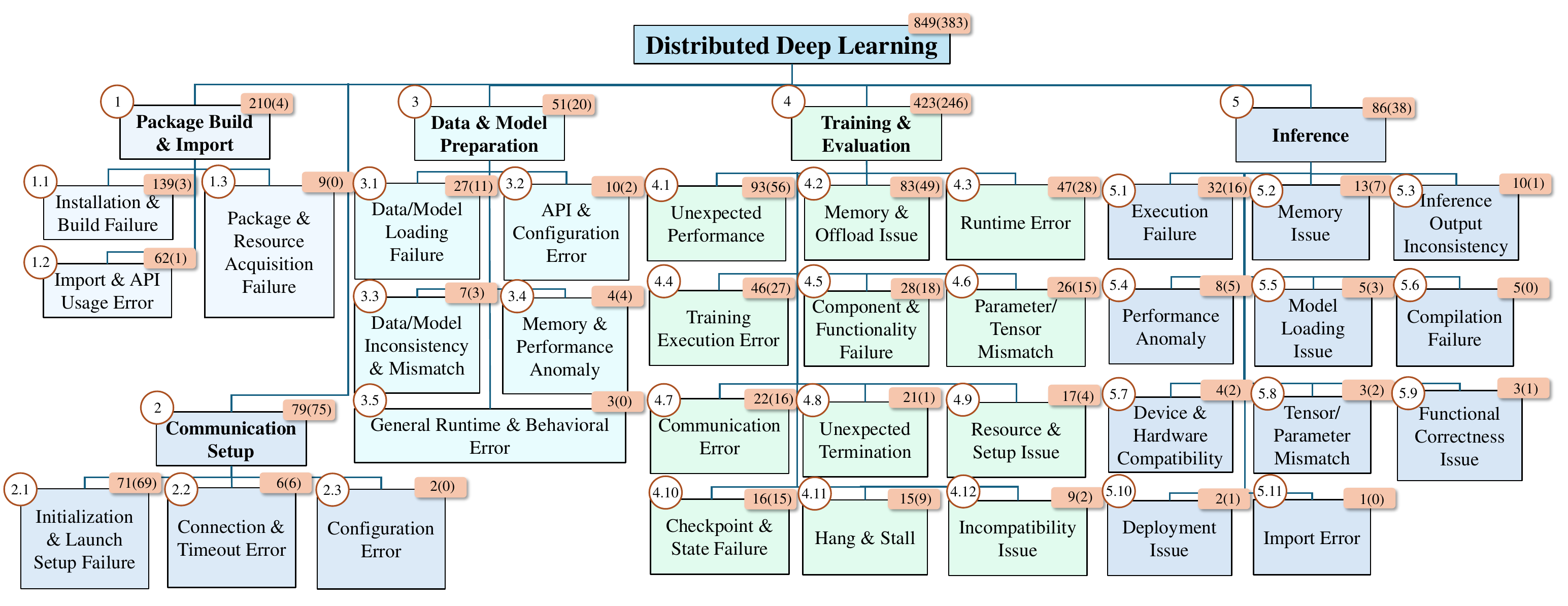}
   \vspace{-0.8cm}
   \caption{Taxonomy of symptoms in distributed training and inference.}
   \label{fig:stages}
   \vspace{-0.4cm}
\end{figure}

We construct a symptom taxonomy aligned with the programming steps of distributed deep learning \cite{liu2023rise}. This taxonomy encompasses 34 core bug symptoms specific to distributed deep learning (see Figure~\ref{fig:stages}). 
It is organized into five main branches, with four representing distinct stages of distributed training and one dedicated to inference. Each leaf node in the taxonomy denotes a specific bug category, and displays two numbers in its upper right corner: the first indicates the total number of issues assigned to the category across all studied frameworks (noting potential overlap with general frameworks like PyTorch and TensorFlow), while the second, shown in parentheses, represents the subset of issues unique to distributed deep learning systems.

\textbf{S.1 Package Build \& Import.} 
Prior to program development, practitioners need to install the distributed deep learning frameworks (e.g., \texttt{pip install colossalai}) and import related modules (e.g., triton). 
Issues occurring at this stage account for 24.7\% of all distributed-related issues, 
with three primary subcategories defining the issues encountered.
\textit{Installation \& build failure (1.1)} constitutes 66.2\% of bugs in this stage and arises during framework installation or source-based building. 
Its most common types are compilation failures (43.9\%), 
such as failures in building pre-compiled operations, and compatibility failures (18.0\%), 
e.g., program crashes due to CPU instruction set incompatibility.
Other issues in this category, ordered by occurrence frequency, include extension failures, build/runtime environment problems, command-not-found errors, missing system headers/dependencies/symbolic links, dependency version conflicts, and miscellaneous issues. 
Following successful framework installation, practitioners frequently encounter \textit{import \& API usage error (1.2)}, which make up 29.5\% of bugs in this stage. 
These errors stem from interactions between code and installed framework components, including module-not-found errors (41.9\%), 
framework version/API incompatibilities (29.0\%), 
incorrect module paths (11.3\%), 
and syntax errors (8.1\%). 
Less common are \textit{package \& resource acquisition failure (1.3)}, representing 4.3\% of bugs in the stage. 
These involve issues in obtaining or initializing resources, such as Docker image unavailability, inability to locate source code, and failures in accessing example scripts or datasets.
In short, only a small subset of issues, such as AVX512-related CPU incompatibility (\textit{1.1}) and framework-specific API incompatibility (\textit{1.2}), are unique to distributed deep learning frameworks, while most problems are also seen in general frameworks.

\textbf{S.2 Communication Setup.}
In distributed deep learning, the establishment of device communication prior to training is a critical prerequisite, particularly when deploying data or model parallelism across multi-GPU or multi-node configurations. Approximately 9.3\% of all issues in distributed workflows arise at this stage, with the majority (94.9\%) 
being exclusive to distributed frameworks rather than general frameworks. 
The issues in this stage can be categorized into \textit{initialization \& launch setup failure (2.1)}, \textit{connection \& timeout error (2.2)}, and \textit{configuration error (2.3)}, which account for 89.9\%, 7.6\%, and 2.5\% of the total, respectively.

Within the initialization and launch setup failure category, 97.2\% of issues are exclusive to distributed frameworks and arise from several distinct sources: misconfiguration of specialized launchers (21.7\%; for example, using \texttt{colossalai run} incorrectly), incompatible network or system configurations (18.8\%; such as protocol mismatches), as well as misconfigurations of multi-node environment managers (15.9\%; including improper SLURM job scheduler setup that disrupts inter-node communication and incorrect node rank allocation with tools like PDSH, which can prevent seamless coordination across compute nodes). Device detection failures are also common (13.0\%; e.g., inability to recognize GPUs in multi-device clusters), as are initialization hangs (11.6\%) linked to distributed communication protocols like MPI and CUDA-aware communication. Additional issues include address or port conflicts, incorrect distributed initialization (leading to problems in inter-process communication and resource allocation), and local rank conflicts (for example, synchronization errors when using DeepSpeed with PyTorch Lightning on multi-node, multi-GPU setups). These cases illustrate how distributed frameworks face unique challenges in initialization and coordination that are rarely seen in traditional single-device environments.

The \textit{NCCL connection \& timeout error (2.2)} subcategory includes failures in inter-device or inter-node communication: multi-GPU connection refusal accounts for 50.0\% of cases (e.g., when GPUs fail to establish basic communication links), NCCL connection failures make up 16.7\% (often triggered by misconfigured firewalls that block NCCL traffic), and NCCL timeouts represent 33.3\% (for instance, when ColossalChat successfully runs on a single GPU but NCCL operations time out during multi-GPU execution). As NCCL is a specialized library for optimizing collective communication in multi-GPU systems, such as all-reduce and broadcast, these errors are unique to distributed deep learning and absent in single-device frameworks. In contrast, the \textit{configuration error (2.3)} category covers general bugs like optimizer setup or data type mismatch, which can also occur in traditional deep learning environments.


\textbf{S.3 Data \& Model Preparation.}
After communication is established, loading data and models to begin training is the next critical operation in distributed deep learning. Here, the use of data, model, or pipeline parallelism leads to failure modes not seen in single-device frameworks. Data and model preparation accounts for 6.0\% of all observed issues, with 39.2\% of those unique to distributed settings. The majority of failures in this stage (52.9\%) are \textit{data/model loading failure (3.1)}, typically linked to model parallelism, examples include mismatches in distributed data partitioning (such as uneven data splits across devices) and errors from distributed API-specific operations (like loading model weights in ColossalAI’s DreamBooth integration). General loading failures, such as missing files (``File Not Found'') or unresolved attributes (``Attribute Not Found''), are also observed. The \textit{API \& configuration error (3.2)} subcategory (19.6\%) includes both distributed-specific problems, such as incompatible distributed feature settings or misconfigured distributed optimizer APIs (e.g., incorrect parameters), and routine mistakes like improper API usage or misconfigured model settings. \textit{Data/Model inconsistency \& mismatch (3.3)} (13.7\%) occurs about equally in distributed and general frameworks, covering distributed issues like failures in meta tensor handling across CPU partitions (e.g., misaligned tensor metadata), as well as more routine issues such as attribute errors or input/model shape mismatches. All cases of \textit{memory \& performance anomaly (3.4)} (7.8\%) are exclusive to distributed frameworks, including CUDA OOM from improper model sharding or increased peak GPU memory use due to suboptimal ZeRO optimization. Finally, \textit{general runtime and behavioral errors (3.5)} (5.9\%) include undefined variable references and anomalies such as incorrect data directory paths or software incompatibilities. These issues occur in both distributed and general-purpose deep learning frameworks.

\textbf{S.4 Training \& Evaluation.} 
Training, where model parameters are iteratively updated through data-driven optimization, is central to deep learning, while evaluation assesses accuracy and generalization on validation data without modifying the trained model. 
This stage is the most critical in deep learning and represents the largest category in our taxonomy, accounting for 49.8\% of identified issues (246 out of 423 total issues). It is further divided into 12 subcategories (labeled 4.1 to 4.12), with the following distribution of bugs: unexpected performance (22.0\%), memory \& offload issue (19.6\%), runtime error (11.1\%), training execution error (10.9\%), component \& functionality failure (6.6\%), parameter/tensor mismatch (6.1\%), communication error (5.2\%), unexpected termination (5.0\%), resource \& setup issue (4.0\%), checkpoint \& state failure (3.8\%), hang \& stall (4.4) (3.5\%), and incompatibility (2.1\%). Notably, over half of the issues (58.2\%) in this stage are unique to distributed deep learning frameworks.

\textit{Unexpected performance (4.1)} encompasses a range of issues that may arise in both general and distributed deep learning frameworks. 
Notably, distributed framework-specific issues are often linked to model or pipeline parallelism. Model parallelism can lead to efficiency degradation (for example, unnecessary memory copying during parameter partitioning in ZeRO-3), numerical instability (such as errors from distributed gradient handling across GPUs, including loss scaling failures or unmitigated gradient overflows), and partially populated tensors resulting from model sharding. 
Pipeline parallelism may trigger unexpected intermediate outputs, such as malformed tensors at various pipeline stages. Additionally, some anomalies, like incorrect logging of model parameters within distributed setups, are unique to distributed environments. In contrast, general frameworks more often encounter issues like learning rate discontinuities and unavailable gradients, which are not tied to parallel training strategies or distributed execution.
For \textit{memory \& offload issue (4.2)}, while some general problems, such as CPU OOM errors and CUDA illegal memory access, can occur in any framework, the majority are specific to distributed frameworks. These include CUDA/GPU memory errors, such as GPU OOM due to ZeRO-3 or unbalanced GPU memory allocation, CPU and system RAM exhaustion, storage and offload-related failures (like inefficient memory offloading or NVMe offload AIO failures), as well as ZeRO-3-related memory leaks and suboptimal memory optimization in distributed environments.
\textit{Runtime error (4.3)} encompasses common issues such as attribute errors, not implemented errors, undefined variable errors, and API usage mistakes, which can arise in both distributed and general deep learning frameworks. 
\textit{Training execution error (4.4)} includes anomalies or instabilities during training, with distributed-specific examples such as ZeRO-3 multi-model configuration errors and training interruptions resulting from improper loss scaling when using ZeRO with quantization.

\textit{Component \& functionality failure (4.5)} refers to the malfunction or misbehavior of core framework elements and distributed functionalities, including examples like AIO synchronization errors (for instance, NVMe asynchronous I/O cross-node offload synchronization failures leading to incomplete data transfers) and shared checkpoint problems (such as checkpoint corruption from race conditions in 8-GPU setups without proper locking). 
\textit{Parameter/Tensor mismatch (4.6)} cover general issues like improper tensor shapes or types (e.g., input layers expecting three channels but receiving one) and distributed-specific errors like dimension mismatches in NVMe-offloaded tensor shards across nodes that cause fusion errors.
\textit{Communication error (4.7)} predominantly affect distributed workflows and involve issues such as failed inter-node connections, NCCL timeouts during multi-GPU data transfer, communication hangs, tensor parallel overlap mismatches, and distributed segmentation faults. 
\textit{Unexpected termination (4.8)} includes process termination due to resource overuse or similar constraints. 
\textit{Resource \& setup issue (4.9)} primarily concern device-related problems, such as unrecognized GPUs.
\textit{Checkpoint failure (4.10)} involves distributed-specific scenarios, such as an inability to load checkpoints due to missing shards on certain nodes. 
\textit{Hang \& stall (4.11)} can occur in both distributed and general frameworks, and include partial GPU completion, the creation of zombie processes, or indefinite training freezes. Finally, \textit{incompatibility (4.12)} includes challenges like API mismatches (e.g., when deprecated calls fail due to version conflicts).


\textbf{S.5 Inference.} 
Inference is the deployment phase in which the trained and validated model generates predictions for new data in practical scenarios. It accounts for 10.1\% of all issues, with distributed inference (running models across multiple devices for scalability) making up 44.2\% of these cases.
The most prevalent distributed-specific issues, ordered by frequency with distributed-only examples, are: \textit{execution failure (5.1)} (37.2\%), including multi-device communication stalls, typos in distributed launcher commands, cross-device data inconsistencies (e.g., mismatched input batches across GPUs), or unexpected process termination; \textit{memory issue (5.2)} (15.1\%), such as sudden GPU memory spikes during forward passes (unique to distributed tensor splitting), excessive RAM usage from framework overhead syncing model shards (causing garbage results), or CUDA OOM errors from poorly sharded models (only when splitting across devices); \textit{inference output inconsistency (5.3)} (11.6\%), marked by mismatched results across devices, often due to tensor parallelism (e.g., split model tensors on different GPUs producing divergent calculations); \textit{performance anomaly (5.4)} (9.3\%), like ineffective parallelization where adding devices fails to reduce latency; \textit{model loading issue (5.5)} (5.8\%), including sharded parameter incompatibility, mismatched data partitioning across nodes, or missing model files in shared storage; \textit{device \& hardware compatibility (5.7)} (4.7\%), featuring distributed-exclusive errors like Triton runtime failures in multi-GPU clusters or GPU kernel incompatibilities with multi-GPU setups; \textit{tensor/parameter mismatch (5.8)} (3.5\%), involving misaligned sharded parameters (e.g., split weight tensors with mismatched dimensions causing crashes); and \textit{deployment issue (5.10)} (2.3\%), such as suboptimal multi-node performance. Other issues, including \textit{compilation failure (5.6) (5.8\%)}, \textit{functional correctness issue (5.9)} (3.5\%), and \textit{import error (5.11)} (1.2\%), occur mostly in general frameworks.

\textbf{For RQ1, see Findings F.1 and F.2, as well as Implications I.1 and I.2 in Table~\ref{table:1}}.


\subsection{Root Causes (RQ2)}
\begin{figure}[!ht]
   \centering
   \includegraphics[width=0.60\linewidth]{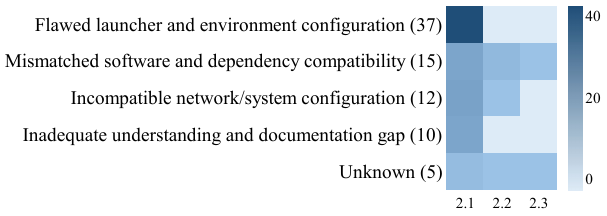}
   \vspace{-0.4cm}
   \caption{Distribution of root causes for categories in communication setup.}
   \label{fig:rc-s2}
\end{figure}

\begin{figure}[!ht]
   \centering
   \includegraphics[width=0.78\linewidth]{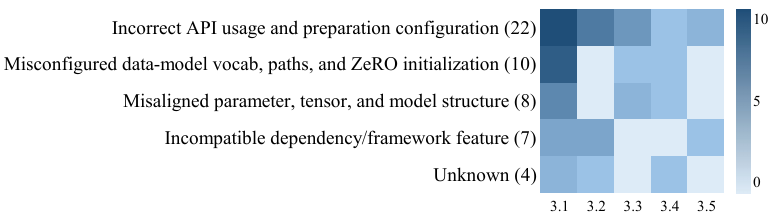}
   \vspace{-0.4cm}
   \caption{Distribution of root causes for categories in data \& model preparation.}
   \label{fig:rc-s3}
\end{figure}

\begin{figure}[!ht]
   \centering
   \includegraphics[width=\linewidth]{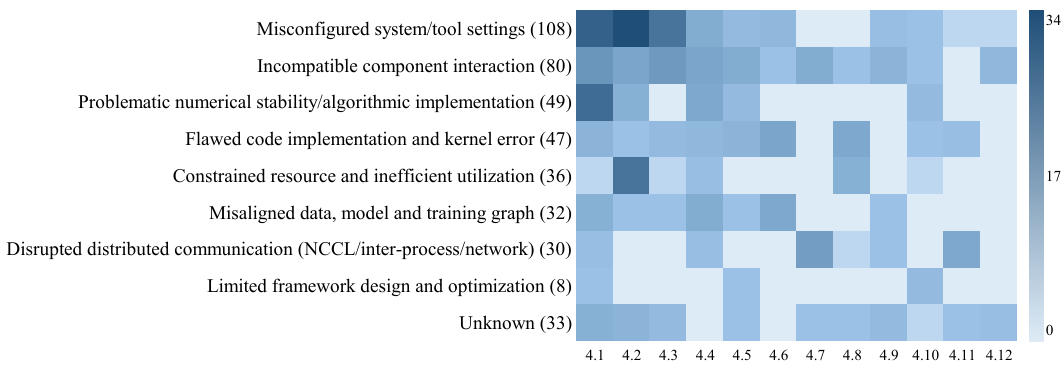}
   \vspace{-0.8cm}
   \caption{Distribution of root causes for categories in training \& evaluation.}
   \label{fig:rc-s4}
\end{figure}

\begin{figure}[!ht]
   \centering
   \includegraphics[width=0.83\linewidth]{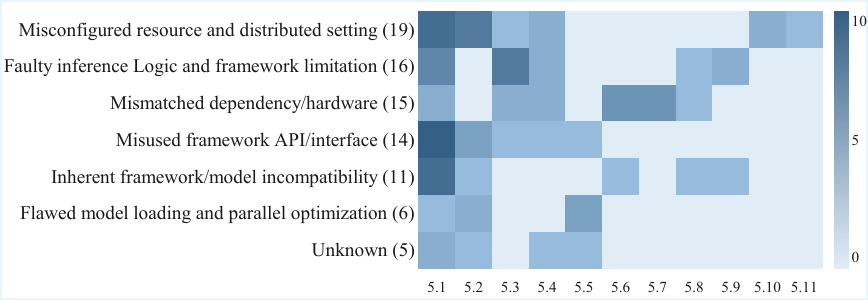}
   \caption{Distribution of root causes for categories in inference.}
   \label{fig:rc-s5}
\end{figure}

For the five stages of distributed deep learning, we counted the frequency of different root causes of issues to explore why bugs occur, as shown in Figures~\ref{fig:rc-s1} to \ref{fig:rc-s5}, respectively. In each figure, the X-axis represents the symptom categories detailed in Section~\ref{sec:sym}, and the Y-axis denotes the root cause categories along with their frequencies for the corresponding stage. We use heatmaps to visualize frequency values, where darker colors indicate higher frequencies. The \textit{Unknown} category includes issues with unidentifiable root causes and is consistently present across all stages.

\subsubsection{Issues in Package Build \& Import}
\begin{figure}[!ht]
   \centering
   \includegraphics[width=0.43\linewidth]{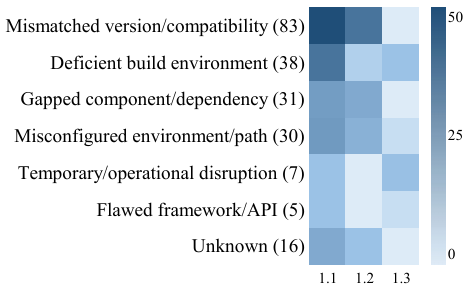}
   \vspace{-0.4cm}
   \caption{Distribution of root causes for categories in package build \& import.}
   \label{fig:rc-s1}
\end{figure}

We identify six root causes in stage 1
(Figure~\ref{fig:rc-s1}). 

\textbf{Mismatched version/compatibility (39.5\%).} 39.5\% of issues in the stage stem from various incompatibilities or incorrect specifications, including mismatched OS versions, conflicting Python versions, inconsistent software/system library versions, incompatible PyTorch APIs, outdated PyPI packages, or incorrect project requirements. 

\textbf{Deficient build environment (18.6\%).} This root cause impacts every symptom category in the stage. Common causes include incorrect configuration of build tools (e.g., wrong CMake paths, mismatched NVCC/GCC versions), missing build tool dependencies (e.g., a required library for GCC linking) triggers ``dependency not found'' errors and cause installation failures, failed Python wheel build process (e.g., corrupted files when packaging a Python wheel) result in failed package installation, and JIT compilation cache problems (e.g., outdated cached data conflicting with new code) that lead to unexpected runtime crashes or import error.

\textbf{Gapped component/dependency (14.8\%).} 
This root cause category includes uninstalled dependencies (e.g., numpy for PyTorch, due to incomplete requirements.txt or offline setups, causing stalls or ``ModuleNotFoundError''), missing system dependencies, unbuilt CUDA extensions, and missing symbolic links (e.g., link cuda pointing to cuda-12.2, preventing tools from locating files).

\textbf{Misconfigured environment/path (14.3\%).} 
Some issues trace back to a misconfigured environment or path problems. Examples include missing CUDA path that causes ``CUDA not found'' errors, incorrect Python env activation, module structure changes that require updated import paths, and wrong Docker command syntax (e.g., should use a Docker image including DeepSpeed v0.6.0 and the patched coop groups files (DeepSpeed issue \texttt{\#1816}\cite{deepspeed_issue1816})).

\textbf{Temporary/operational disruption (3.3\%).}
A few issues are attributed to operational disruptions. These disruptions include problems like website or network connectivity issues that result in resource acquisition failures (for example, an invalid Slack link), incorrect Docker image tagging (such as the inability to pull the latest Docker image in ColossalAI issue \texttt{\#1756}\cite{colossalAI_issue1756}), and the use of outdated compiler versions.

\textbf{Flawed framework/API (1.9\%).} 
A small proportion of issues arise from internal path resolution bugs within the framework, which can lead to copy failures, as well as platform-specific implementation flaws.

\subsubsection{Issues in Communication Setup.}
We identify four root causes in stage 2
(Figure~\ref{fig:rc-s2}).

\textbf{Flawed launcher and environment configuration (46.8\%).}
This is the most common root cause in this stage, specifically related to data/model loading failures. The primary one is improper usage of the distributed launcher or incorrect launch configurations. For example, in DeepSpeed issue \texttt{\#1923}\cite{deepspeed_issue1923}, an incorrect multi-node launcher invocation leads to failures in setting up the DeepSpeed launcher and process group. Other contributing factors include environment variables not being properly propagated (such as CUDA-related variables not being correctly shared across nodes via PDSH, resulting in GPU detection failures), launcher parsing errors, improper environment variable configurations, and misunderstandings regarding the integration of the distributed launcher and environment variables (as seen in Colossal-AI issue \texttt{\#1082}\cite{colossalAI_issue1082}). 

\textbf{Mismatched software and dependency compatibility (19.0\%).}
Some bugs arise from platform incompatibility, such as the failure to run NCCL on Windows (Colossal-AI issue \texttt{\#2941}\cite{colossalAI_issue2941}), as well as compiler incompatibility in distributed builds, version mismatches between frameworks and libraries, dependency version discrepancies, and incompatibility with partially frozen models (Colossal-AI issue \texttt{\#5290}\cite{colossalAI_issue5290}). 

\textbf{Incompatible network/system configuration (15.2\%).}
Some issues are caused by improper SSH setups, as seen in DeepSpeed issue \texttt{\#4034}\cite{deepspeed_issue4034}, where a Docker Swarm cluster fails to execute multi-node operations due to a lack of access between cluster nodes for the DeepSpeed launcher. Other factors include misconfigured network or system environments, such as failures of RLHF Training Stage 1 in ColossalChat when scaling from a single GPU to two, and problems with firewall or network configuration that impede NCCL operations. 

\textbf{Inadequate understanding and documentation gap (12.7\%).}
Operational misunderstandings and insufficient documentation contribute to numerous issues. For instance, DeepSpeed issue \texttt{\#5278}\cite{deepspeed_issue5278} involves a launcher failing to convert an environment variable to an integer due to an incorrect assumption (i.e., the variable only accepts integers) about accepted formats, which conflicts with documentation allowing UUIDs. 
Similarly, lack of familiarity with parameter substitution syntax in PDSH leads to incorrect configuration of node rank in DeepSpeed issue \texttt{\#1523}\cite{deepspeed_issue1523}. 
As commented by a DeepSpeed collaborator, ``This is a bit subtle and should definitely be better documented here since it's not obvious when reading the code''. 

\subsubsection{Issues in Data \& Model Preparation.}
We identify four root causes in stage 3
(Figure~\ref{fig:rc-s3}).

\textbf{Incorrect API usage and preparation configuration (43.1\%).}
The primary cause of failures when preparing data or models is incorrect use of framework APIs or misconfigurations, which can lead to assertion errors or memory issues. For example, in DeepSpeed issue \texttt{\#2672}\cite{deepspeed_issue2672}, a user encountered a CUDA OOM error when loading the facebook/opt-66b model onto up to 96 A100-80 GPUs using ZeRO-3, due to suboptimal model sharding configurations during initialization.

\textbf{Misconfigured data-model vocab, paths, and ZeRO initialization (19.6\%).}
One-fifth of issues stem from misconfigured data paths/structures, data-model vocab mismatch, or flawed ZeRO Stage model initialization. 
For example, Colossal-AI issue \texttt{\#2605}\cite{colossalAI_issue2605} saw a user hit the error ``PytorchStreamReader failed reading zip archive: failed finding central directory'' due to a misconfigured data path pointing to a corrupted/unreachable file. In DeepSpeed issue \texttt{\#3734}\cite{deepspeed_issue3734}, improper ZeRO Stage setup (e.g., wrong ZeRO Stage 3 loading strategies) triggered transient memory spikes.

\textbf{Misaligned parameter, tensor, and model structure (15.7\%).}
Some issues result from misaligned parameters, tensors, or model structures, including mismatched parameters/tensors, incompatible parameter slicing, and inappropriate model parameter configurations. For example, in DeepSpeed issue \texttt{\#1770}\cite{deepspeed_issue1770}, a user could not initialize the inference engine with GPT2 due to a tensor type mismatch in internal copy logic (fixed by upgrading). Colossal-AI issue \texttt{\#1402}\cite{colossalAI_issue1402} involved a pre-trained Pegasus model loading failure: old ZeRO sliced parameters into 1D tensors, breaking shape checks during initialization.

\textbf{Incompatible dependency/framework feature (13.7\%).}
A few issues are caused by mismatched dependency versions or incompatible framework features/versions. For example, in Colossal-AI issue \texttt{\#3202}\cite{colossalAI_issue3202}, it was reported that the FusedSGD optimizer was incompatible with the ZeRO mechanism, leading to a configuration error. Additionally, an update in Stable Diffusion v2, specifically a change in the source code group size from 32 to 16, introduced a potential configuration incompatibility (Colossal-AI issue \texttt{\#3109}\cite{colossalAI_issue3109}).

\subsubsection{Issues in Training \& Evaluation.}
We identify eight root causes in stage 4
(Figure~\ref{fig:rc-s4}).

\textbf{Misconfigured system/tool setting (25.5\%).}
This category includes errors from incorrect or incomplete system configurations, improper model sharding settings, misuse of distributed launchers, and faulty ZeRO-3 configurations. For instance, if all GPUs, including low-memory devices, are used without explicit exclusion (DeepSpeed issue \texttt{\#3235}\cite{deepspeed_issue3235}), CUDA OOM errors can arise. In DeepSpeed issue \texttt{\#2608}\cite{deepspeed_issue2608}, misconfiguring the \texttt{redundancy\_clean} API in ZeroQuant compression prevented the expected size reduction, resulting in inefficient performance. ColossalAI issue \texttt{\#2145}\cite{colossalAI_issue2145} involved resource misallocation and memory mismanagement for mT5 model kernels, leading to tensor shape and type errors during parallel computation.

\textbf{Incompatible component interaction (18.9\%).}
This category encompasses mismatches between hardware, software, and frameworks during distributed deep learning training and evaluation. Typical issues include hardware incompatibility, conflicting dependency versions, model-framework mismatches, inconsistent distributed checkpoints or APIs, and limited support for distributed optimization features. 
For example, if ColossalAI or a required dependency is not fully compatible with the active Python version in a Conda environment, running \texttt{torchrun} may start training but result in runtime failures due to version discrepancies between PyTorch and CUDA, such as distributed CUDA OOM errors.
DeepSpeed issue \texttt{\#4648}\cite{deepspeed_issue4648} illustrated a hardware or kernel mismatch in which the GPU, its drivers, or firmware failed to meet the requirements of the CUDA or NCCL libraries. This resulted in resource-related errors, such as device failures during distributed training.


\textbf{Problematic numerical stability/algorithmic implementation (11.6\%).}
The category involves errors from unstable numerical computation or flawed algorithms in distributed deep learning. Main contributors are misconfigured mixed precision, FP16 instability, distributed kernel inaccuracies, and algorithmic misunderstandings.
For example, ColossalAI issue \texttt{\#1035}\cite{colossalAI_issue1035} was due to an incorrect mixed precision scaling (i.e., PyTorch Automatic Mixed Precision’s interaction with gradient accumulation skips early parameter updates), causing execution errors in multi-model (ZeRO-3) setups. DeepSpeed issue \texttt{\#1356}\cite{deepspeed_issue1356} exposed FP16 precision limitations, where the framework failed to revert to higher precision operations when needed, resulting in numerical instability and poor convergence.


\textbf{Flawed code implementation and kernel error (11.1\%).}
Some issues result from programming mistakes and kernel bugs.
These include in-place changes during communication, parameter mismatches, tensor handling problems, timer or kernel bugs, optimizer and logging tool errors, high-load data failures, random number mismanagement, and storage bugs.
For example, ColossalAI issue \texttt{\#4412}\cite{colossalAI_issue4412} was triggered by high-load data processing that omitted the required \texttt{attention\_mask} key in training batches, causing ZeRO-3 execution errors. In ColossalAI issue \texttt{\#1051}\cite{colossalAI_issue1051}, manual seed settings conflicted with internal RNG state management and led to unexpected termination by the operating system.


\textbf{Constrained resource and inefficient utilization (8.5\%).}
This category includes issues where hardware or memory limits are reached, or resources are not used effectively. The main types include memory issues (e.g., OOM errors, memory leaks, inefficient memory management, and boundary violations), hardware/storage shortages, inefficient data/operation handling, and checkpointing/load problems (e.g., not enough shared memory or progressive memory leaks during checkpoint saving/loading).
For example, in DeepSpeed issue \texttt{\#3340}\cite{deepspeed_issue3340}, a user set a large batch size for the Bloom-1.1b model that exceeded V100 GPU memory, resulting in resource exhaustion. In DeepSpeed issue \texttt{\#2632}\cite{deepspeed_issue2632}, system memory runs out, triggering the OOM killer and abruptly ending the training process unexpectedly.


\textbf{Misaligned data, model, and training graph (7.6\%).}
This category contains alignment errors among data, model architecture, and the computational graph, such as inconsistent input-weight types, distributed loading problems, incompatible vocabularies, parameters or architectures unsuited for parallelization, and training graph flaws.
For example, in DeepSpeed issue \texttt{\#4829}\cite{deepspeed_issue4829}, a user called the GPT-J-6B model twice in serial order during forward propagation, causing DeepSpeed to process gradient calculations as if training two independent models. This led to duplicated computation costs and a CUDA OOM error. 


\textbf{Disrupted distributed communication (NCCL/inter-process/network) (7.1\%).}
This category includes failures in NCCL P2P communication, workload imbalance across processes, synchronization errors in tensor and pipeline parallelism, network or system configuration incompatibilities, and host-level communication issues within distributed clusters. For example, DeepSpeed issue \texttt{\#2223}\cite{deepspeed_issue2223} was caused by uneven GPU process batch counts, disrupting inter-process synchronization and resulting in a training stall.

\textbf{Limited framework design and optimization (1.9\%).} 
This root cause refers to failures stemming from inadequate framework architecture or suboptimal implementation of optimization strategies. It covers design limitations in the framework, improper tool usage, and flawed implementations of distributed communication mechanisms.
For example, ColossalAI issue \texttt{\#990}\cite{colossalAI_issue990} stemmed from incompatibility with PyTorch 1.9.0+cu111, leading to segmentation faults and program crashes during distributed operations.

\begin{table}[!ht]
\vspace{-0.2cm}
\centering
\scriptsize
\renewcommand{\arraystretch}{0.6}
\caption{Summary of Findings and Implications}
\label{table:1}
\vspace{-0.4cm}
\begin{tabular}{p{5cm} | p{8cm}}
\toprule
\textbf{Findings about symptoms}       & \textbf{Implications} \\\midrule
\textbf{F.1.} We construct a taxonomy of 34 symptoms for bugs in distributed deep learning, with 45.1\% occurring exclusively in distributed frameworks. 
&     \textbf{I.1.} 
The diversity and framework-specific nature of distributed deep learning bugs—nearly half of which are unique to distributed systems—means that framework developers and users must go beyond traditional error checks. Practical steps like implementing stricter API validation and running version compatibility tests before training can help catch subtle config or code mismatches. For example, having the framework automatically check that tensor shapes and data types are aligned across all nodes before distributed training begins allows errors to be flagged clearly in logs, preventing confusing runtime crashes and making debugging much more manageable.
\\\midrule

\textbf{F.2.} Among these exclusive symptoms, the most common are initialization and launcher setup failures, memory issues, and unexpected performance.   &    

\textbf{I.2.} 
To address issues like initialization failures, memory errors, and performance bottlenecks that are common in distributed frameworks, framework developers and system administrators may build automated resource and path checks into setup routines, such as scripts that confirm device connectivity and required files before launching jobs. Users can also improve reliability by using GPU monitoring tools (like nvidia-smi) to adjust batch sizes and prevent OOM crashes. To spot and resolve communication slowdowns, both developers and users may regularly employ profiling tools such as PyTorch Profiler, which help visualize and optimize data partitioning and node communication for smoother, more efficient training.
\\\midrule   



\textbf{Findings about root causes}       & \textbf{Implications} \\\midrule

 \textbf{F.3.} In \textit{communication setup}, 95\% of issues are distributed framework-specific, including multi-node problems like misconfigured launchers, rank errors, and NCCL/multi-GPU failures. They are rooted primarily in poor launcher/env configs, software/dependency conflicts, and network/system incompatibilities (e.g., NCCL settings). &
 
 \textbf{I.3.} Framework developers and support teams should create and maintain clear setup guides with real command-line examples, and build automated scripts to check environment variables and node connectivity before launching jobs. On the user side, taking a moment to run simple SSH or network checks between nodes before starting an experiment can catch tricky errors early. While these steps may seem basic, they go a long way toward preventing frustration and helping everyone, from beginners to experts, avoid some of the frustrating distributed training errors.
        \\\midrule
        
 \textbf{F.4.} Most issues in \textit{data \& model preparation} arise from incorrect API usage, configuration errors, poor ZeRO initialization, and parameter/model mismatches. These cause model parallelism-related failures like CPU meta tensor errors and CUDA OOM/peak memory from improper sharding/ZeRO. & 
 
 \textbf{I.4.} 
Users should follow framework API guidelines closely and double-check all configuration files to avoid subtle mistakes, such as using the wrong ZeRO initialization settings or pointing to incorrect data paths, which can cause memory spikes or model loading errors. It is also important to verify parameter shapes and model structure compatibility before training, as mismatches can trigger failures that are hard to debug later. Finally, regularly updating and checking dependencies helps prevent incompatibility issues when frameworks, models, or optimizers are upgraded or changed.
         \\\midrule
         
 \textbf{F.5.}  In \textit{training \& evaluation}, most issues are caused by misconfiguration and incompatibility in system and tool settings, unstable numerical methods, and software or kernel errors, resulting in frequent crashes, resource exhaustion, and unexpected model performance. &
 
\textbf{I.5.} During distributed training and evaluation, users should determine each GPU’s memory and adjust batch sizes in the job config (e.g., use \texttt{CUDA\_VISIBLE\_DEVICES} to exclude any slow ones). 
It is necessary to review and set mixed-precision and loss scaling settings in the framework (e.g., update the fp16 section of your DeepSpeed JSON config) and monitor logs for any overflow or instability. Additionally, adding assertion checks in key areas of the training code can help verify tensor shapes after partitioning and at various pipeline stages. Explicit resource limits, such as \texttt{max\_memory\_allocated}, should be reviewed for each worker, and logs should be actively monitored for OOM or crash warnings to allow for timely intervention before a process fails. 
        \\\midrule
        
  \textbf{F.6.} Many issues arise from improper model sharding, misconfigurations in memory or inference, flawed internal framework inference logic, or mismatched dependencies and hardware during \textit{inference}. These factors primarily lead to execution failures, such as stalls and unexpected terminations, as well as OOM errors and inconsistent results across devices.  & 

 \textbf{I.6.} For distributed inference, users should select the sharding strategy (such as row-wise or column-wise splits) that best matches the model architecture and hardware, check memory settings to avoid OOM errors, and confirm all nodes use the same dependencies and environment before running predictions. Unlike training, inference prioritizes fast model loading and consistent outputs across devices. Before executing real predictions, it is helpful to send sample queries and use built-in hardware check tools to identify malfunctioning devices or problematic settings. Additionally, verify settings like max token limits and ensure data flows align with model expectations to ensure smooth and reliable inference across all devices. 
  \\\midrule
  
 \textbf{F.7.}
Over 90\% of issues can be resolved using no more than four fix patterns. Managing versions and dependencies, as well as tuning distributed APIs, communication, and features, are the two most prominent strategies, appearing at every stage and accounting for 37\% and 32\% of all issues, respectively. &

 \textbf{I.7.} 
To maximize usability and reliability, framework developers should implement platform-specific checks, provide comprehensive installation guides, and ensure compatibility across all major operating systems. They should also prioritize clear version and dependency management (e.g., synchronizing CUDA versions and automating environment setup with Docker or Conda) to reduce build errors and streamline debugging. Additionally, offering sample configurations, intuitive APIs for various ZeRO stages and multi-model setups, and practical troubleshooting resources will empower users to efficiently resolve integration issues and enhance overall framework adoption and robustness.
 \\\bottomrule                   
\end{tabular}
\end{table}

\subsubsection{Issues in Inference.}
We identify six root causes in stage 5
(Figure~\ref{fig:rc-s5}).

\textbf{Misconfigured resource and distributed setting (22.1\%).} 
Root causes here involve faulty model sharding, insufficient shared memory, incorrect command-line arguments, missing distributed inference settings, parameter mismatches, and improper Python environment setup. 
For example, DeepSpeed issue \texttt{\#2917}\cite{deepspeed_issue2917} arose from wrong model sharding settings: when inferring the 176B Bloom model, DeepSpeed’s sharding and resource management failed to allocate memory across CPU and GPU effectively, exhausting CPU virtual memory and causing a CUDA OOM error. DeepSpeed issue \texttt{\#2558}\cite{deepspeed_issue2558} stemmed from incorrect model parameters. When calling the inference engine, the \texttt{mp\_size} (desired model parallelism) conflicted with the actual distributed world size, failing PyTorch distributed group initialization and triggering an API usage error.


\textbf{Faulty inference Logic and framework limitation (18.6\%).}
This category involves errors due to incorrect batch processing, framework limitations (e.g., lack of support for dynamic sequence operations), kernel token limits, inference generation problems, mismatched sampling behavior, and improper settings for content filtering or generation parameters.  
For example, DeepSpeed’s (\texttt{kernel\_inject}) optimization suffered from internal logic flaws related to \texttt{MAX\_OUT\_TOKES} during long-sequence inference, causing unstable model outputs and inconsistent inference results.


\textbf{Mismatched dependency/hardware (17.4\%).} 
This category includes issues caused by compiler and environment mismatches, dependency version incompatibilities, T4 GPU kernel incompatibilities, and software conflicts such as DeepSpeed-Triton integration problems. For example, DeepSpeed issue \texttt{\#2607}\cite{deepspeed_issue2607} failed to compile due to a mismatch between the PyTorch (g++) and nvcc (nvc++/PGI) compilers, demonstrating how incompatible components can result in build and runtime failures.


\textbf{Misused framework API/interface (16.3\%).}
This category includes failures from incorrect or misunderstood use of framework APIs, such as improper handling of distributed APIs for tensor parallelism and missing model initialization. For example, in DeepSpeed-Inference AutoTP (issue \texttt{\#7278}\cite{deepspeed_issue7278}), a user misused the \texttt{split\_between\_processes} method, resulting in inconsistent inputs across GPU processes and causing mismatched inference outputs.


\textbf{Inherent framework/model incompatibility (12.8\%).}
It covers model, framework, or platform mismatches, limitations in framework design (such as multi-engine tensor parallel size problems), and conflicts between frameworks and deployment tools. 
For example, in DeepSpeed issue \texttt{\#2895}\cite{deepspeed_issue2895}, when processing Bloom models, DeepSpeed’s inference optimization altered internal tensor shapes like \texttt{attention\_mask}, conflicting with the model’s attention mechanism expectations and causing runtime errors during inference.


\textbf{Flawed model loading and parallel optimization (7.0\%).}
This root cause centers on inefficient memory usage during model loading for parallel inference. DeepSpeed issue \texttt{\#1772}\cite{deepspeed_issue1772} showed that the default behavior of the transformers library, combined with parallel multi-process startup, can cause total CPU memory consumption to exceed system limits when working with very large models. As a result, distributed inference may abruptly terminate when available memory is exhausted.

\textbf{For RQ2, see Findings F.3 to F.6, as well as I.3 to I.6 in Table~\ref{table:1}.}


\begin{figure}[!th]
    \centering
    \includegraphics[width=0.32\linewidth]{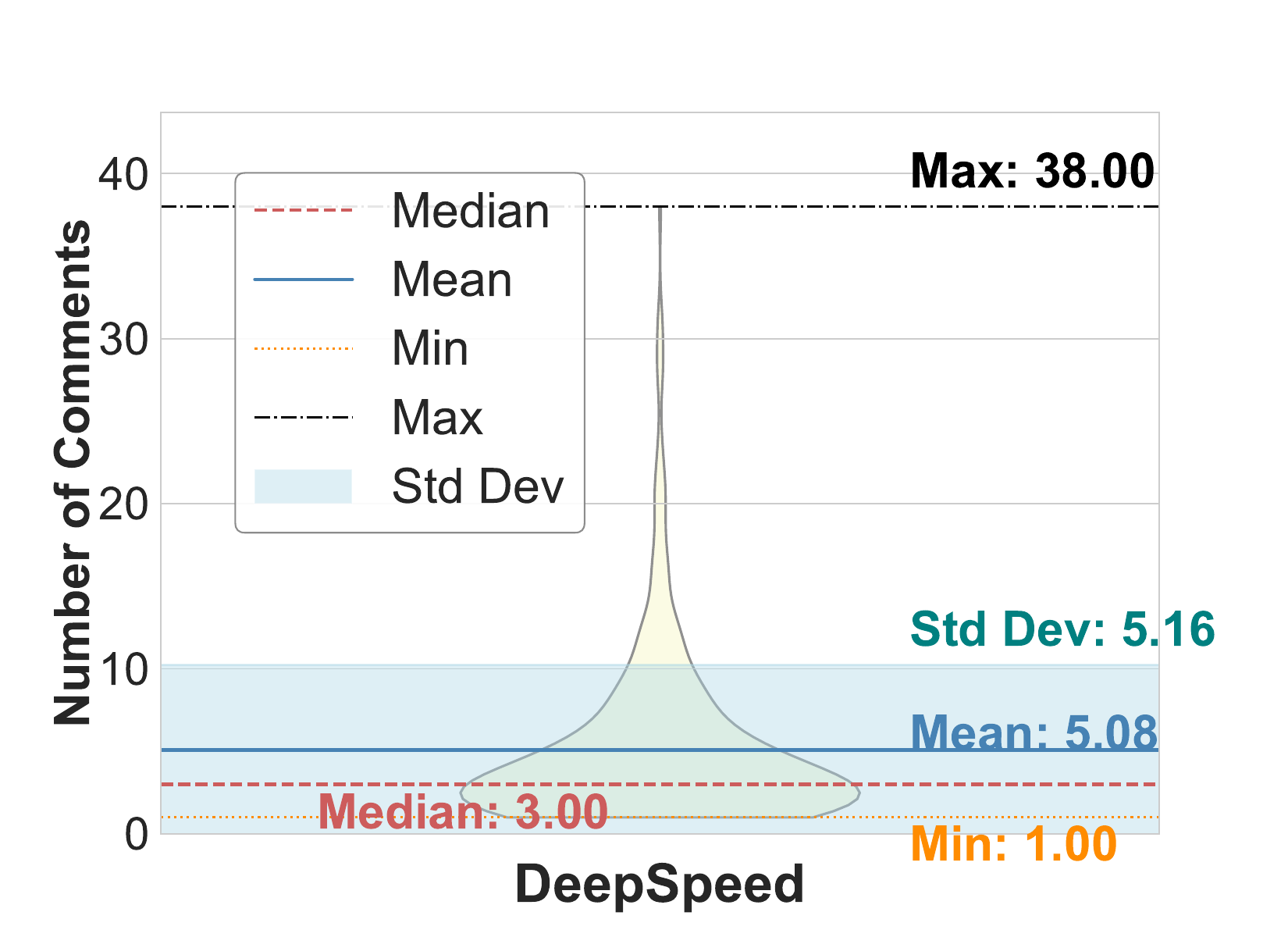}
    \hfill
    \includegraphics[width=0.32\linewidth]
    {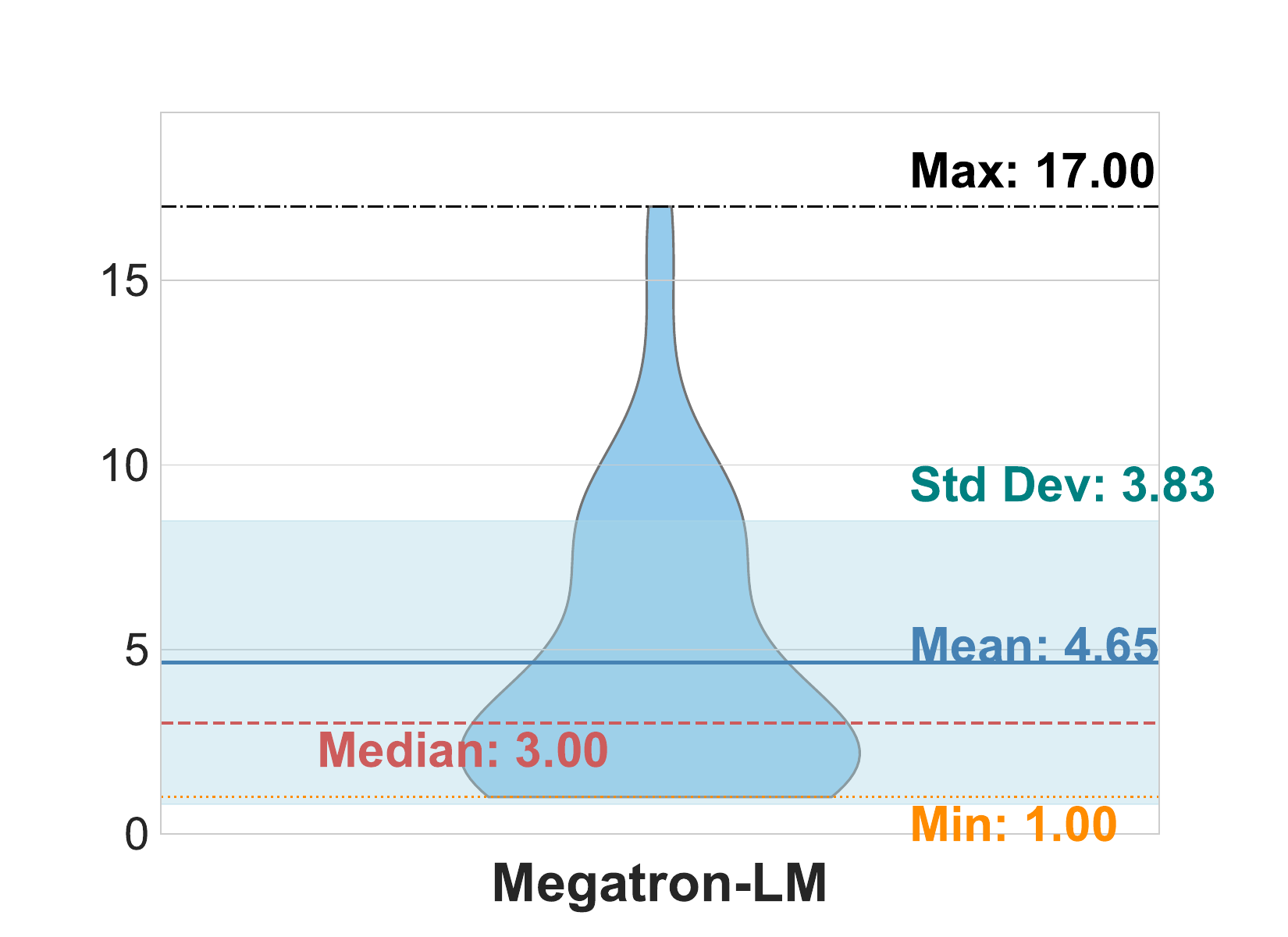}
    \hfill
    \includegraphics[width=0.32\linewidth]
    {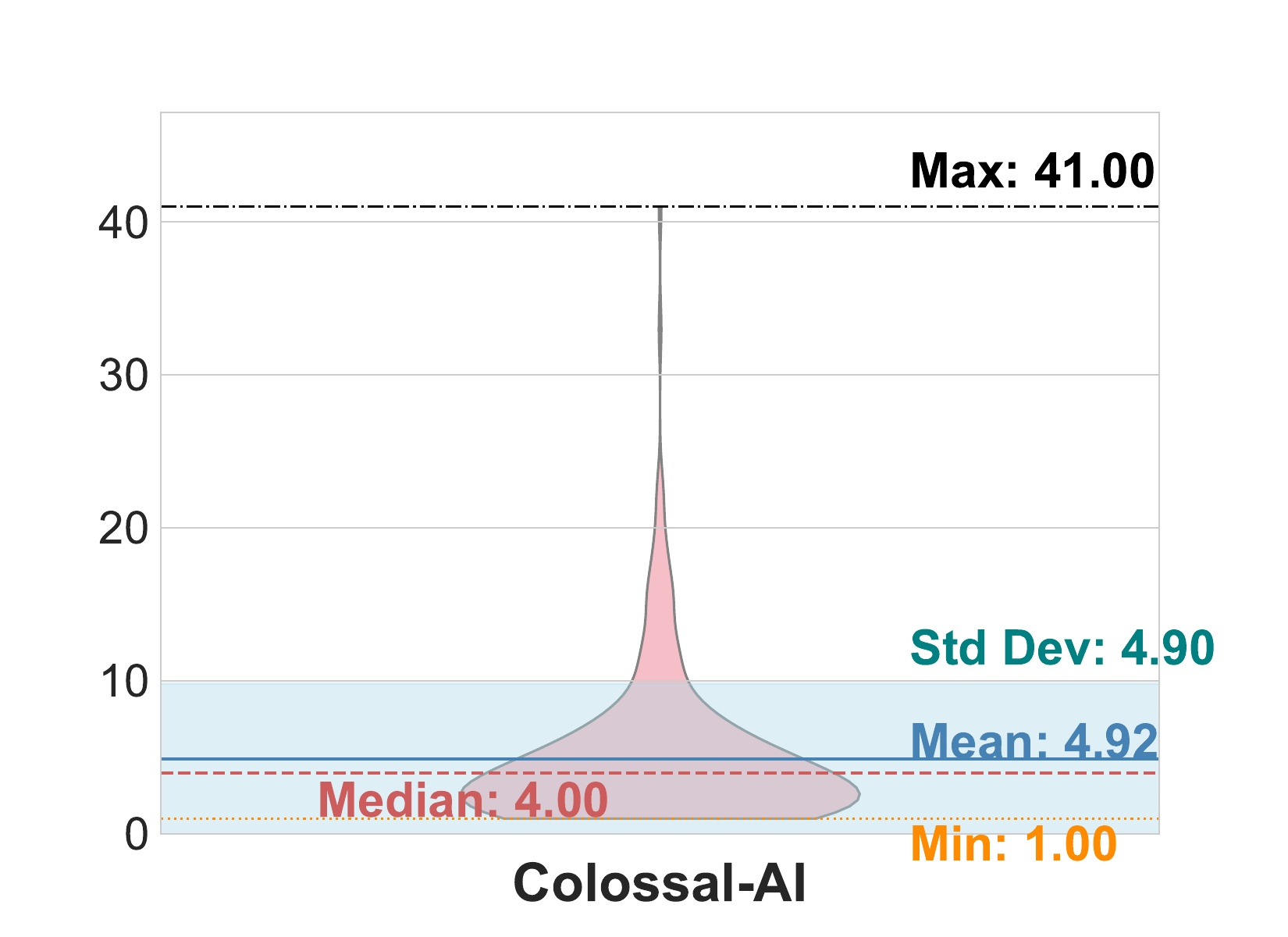}
    \includegraphics[width=0.32\linewidth]{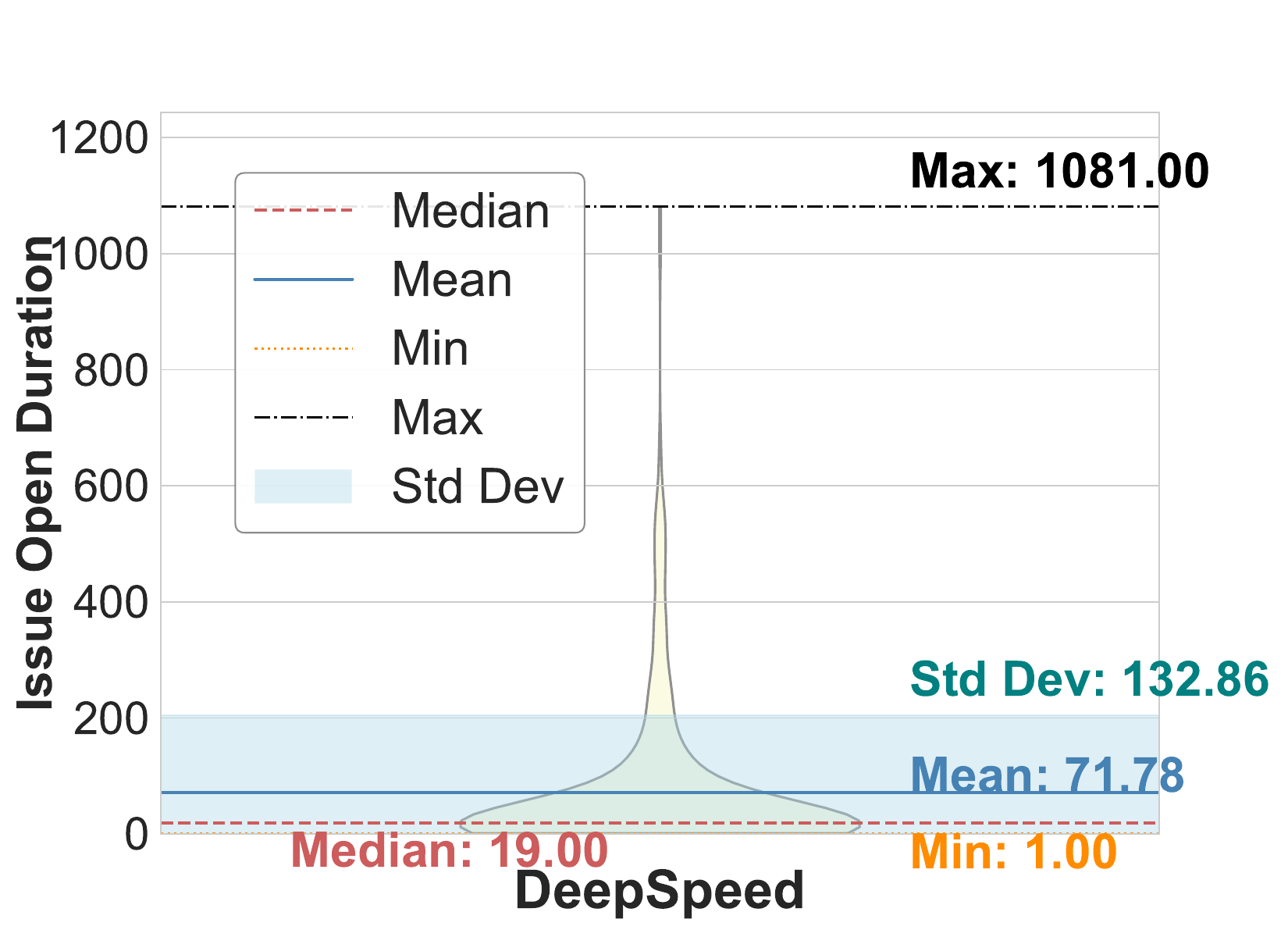}
    \hfill
    \includegraphics[width=0.32\linewidth]{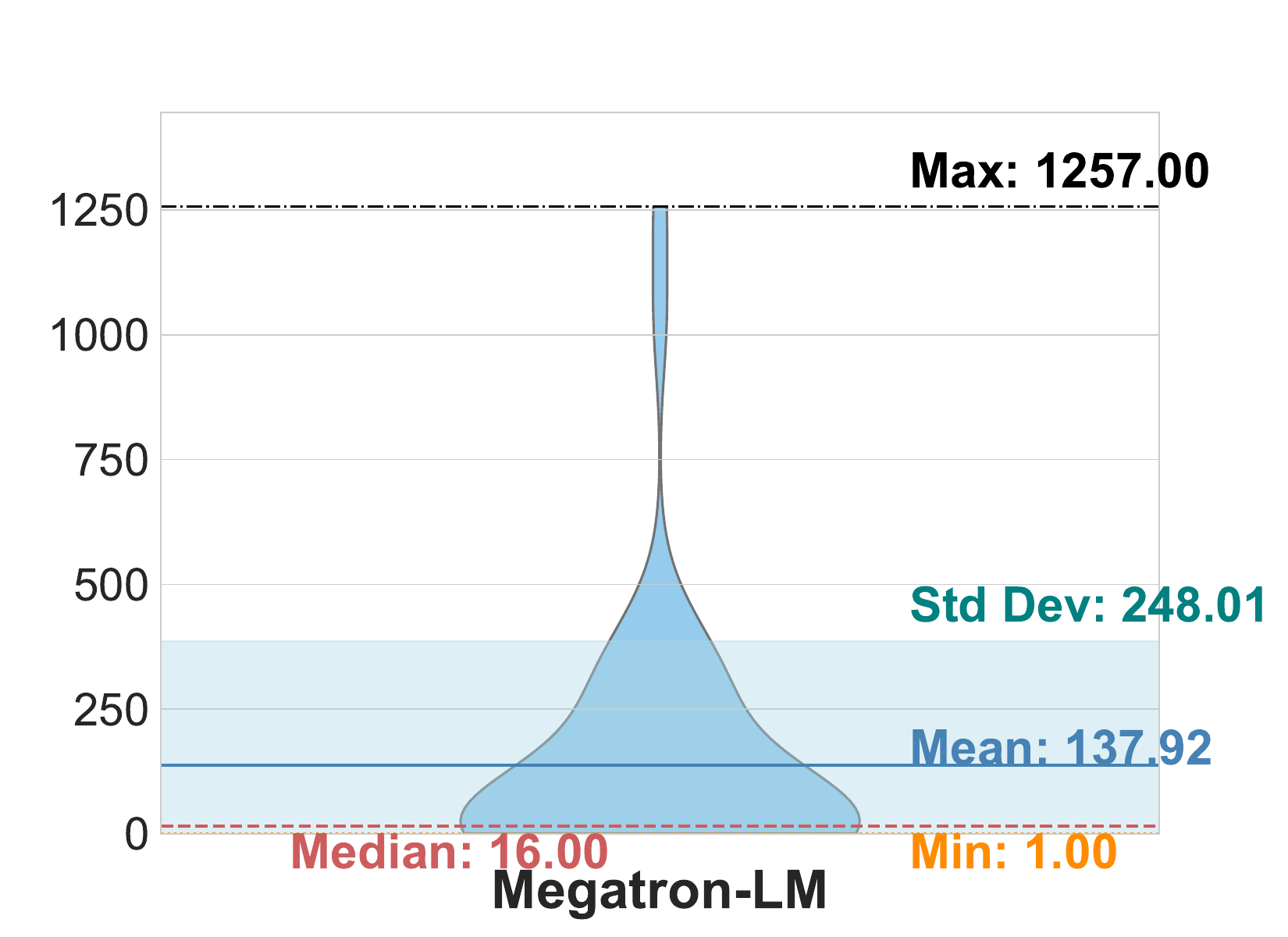}
    \hfill
    \includegraphics[width=0.32\linewidth]{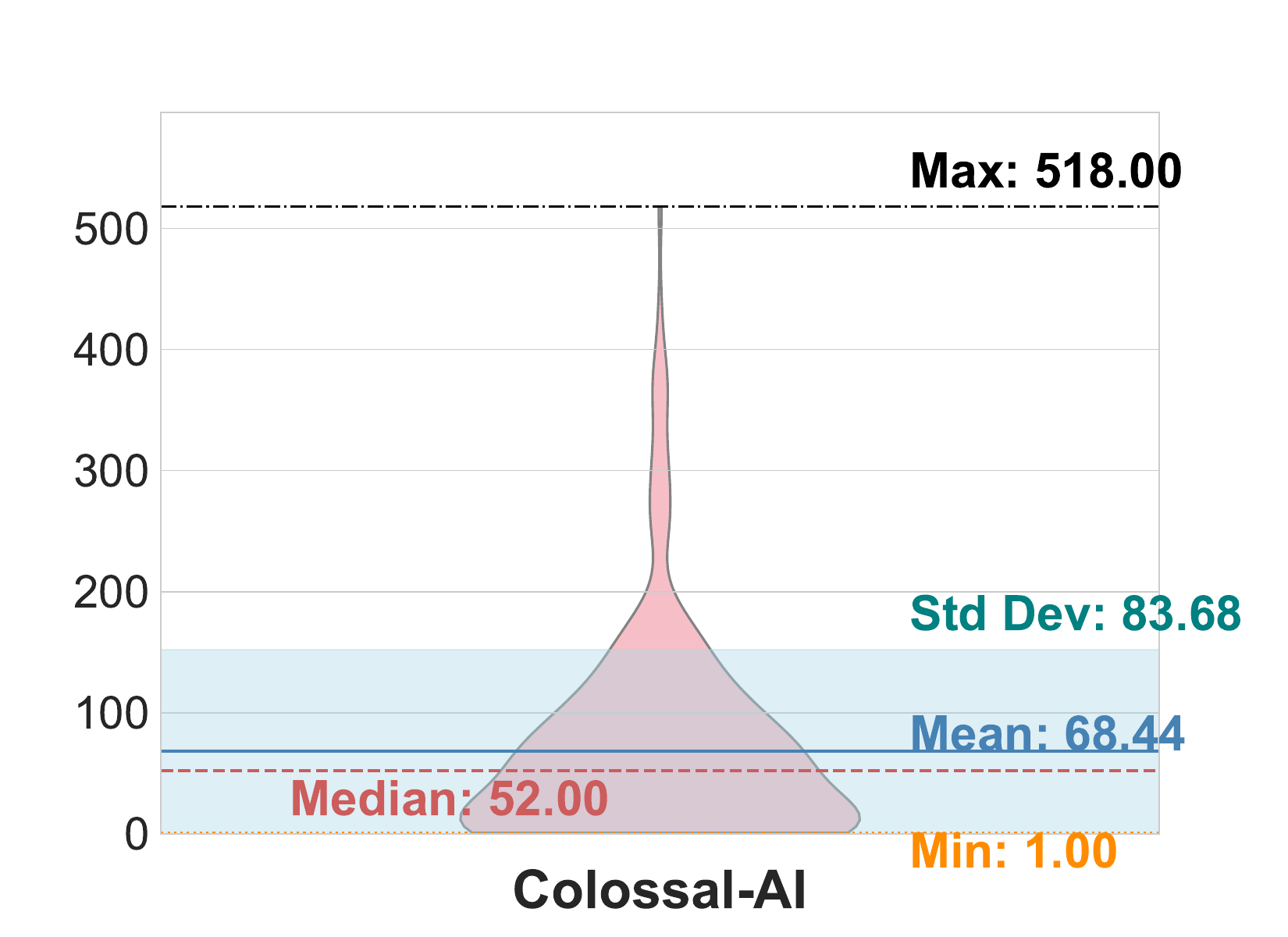}
   \caption{Distribution of the comment counts and open durations for reported issues.}
   \label{fig:violin 1}
\end{figure}

\subsection{Fix Patterns}

We analyzed all issues from their publication to closure, excluding those in the \textit{Unknown} root cause category as these lack identifiable fixes and clear patterns. This subset of excluded issues 
ensures that the presented statistics focus on actionable cases with determinable causes and resolutions. 
Figure~\ref{fig:violin 1} presents violin plots of the number of comments per issue and issue open duration across three frameworks. Megatron-LM shows a lower maximum comment count, likely due to its smaller total number of issues compared to DeepSpeed and Colossal-AI. However, all three frameworks share similar comment statistics, with means, medians, and standard deviations typically ranging from 3 to 5. Single-comment issues are common, often resulting from immediate resolution and closure by the reporter, or from developer responses that go unanswered, leading to issues being marked stale and closed. Regarding open duration, Megatron-LM’s issues tend to remain open longer. This is probably because DeepSpeed and Colossal-AI boast larger user bases and more active development communities, enabling more timely responses and discussions.

\begin{figure}[!th]
    \centering
    \includegraphics[width=0.32\linewidth]{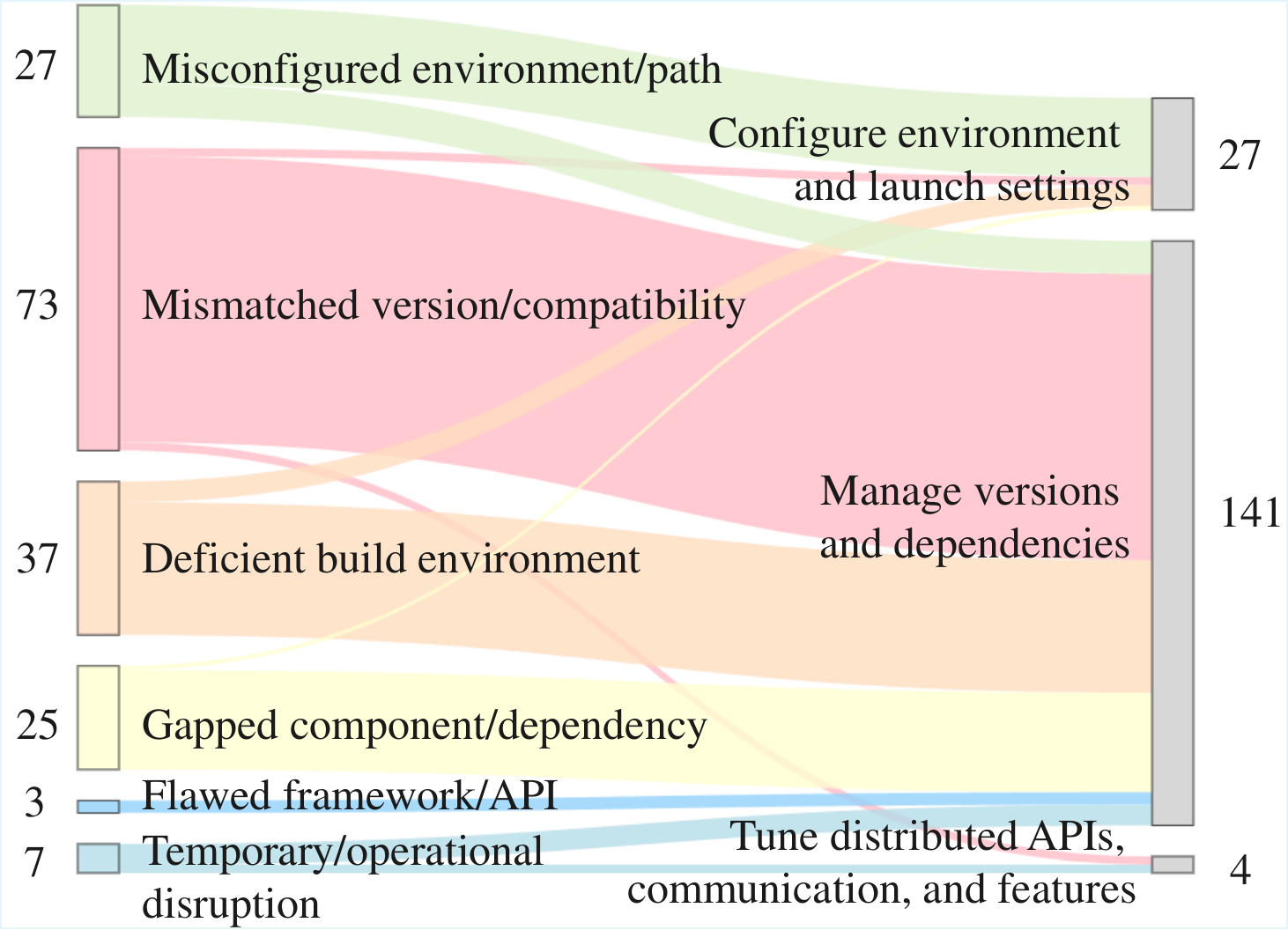}
    \hfill
    \includegraphics[width=0.32\linewidth]{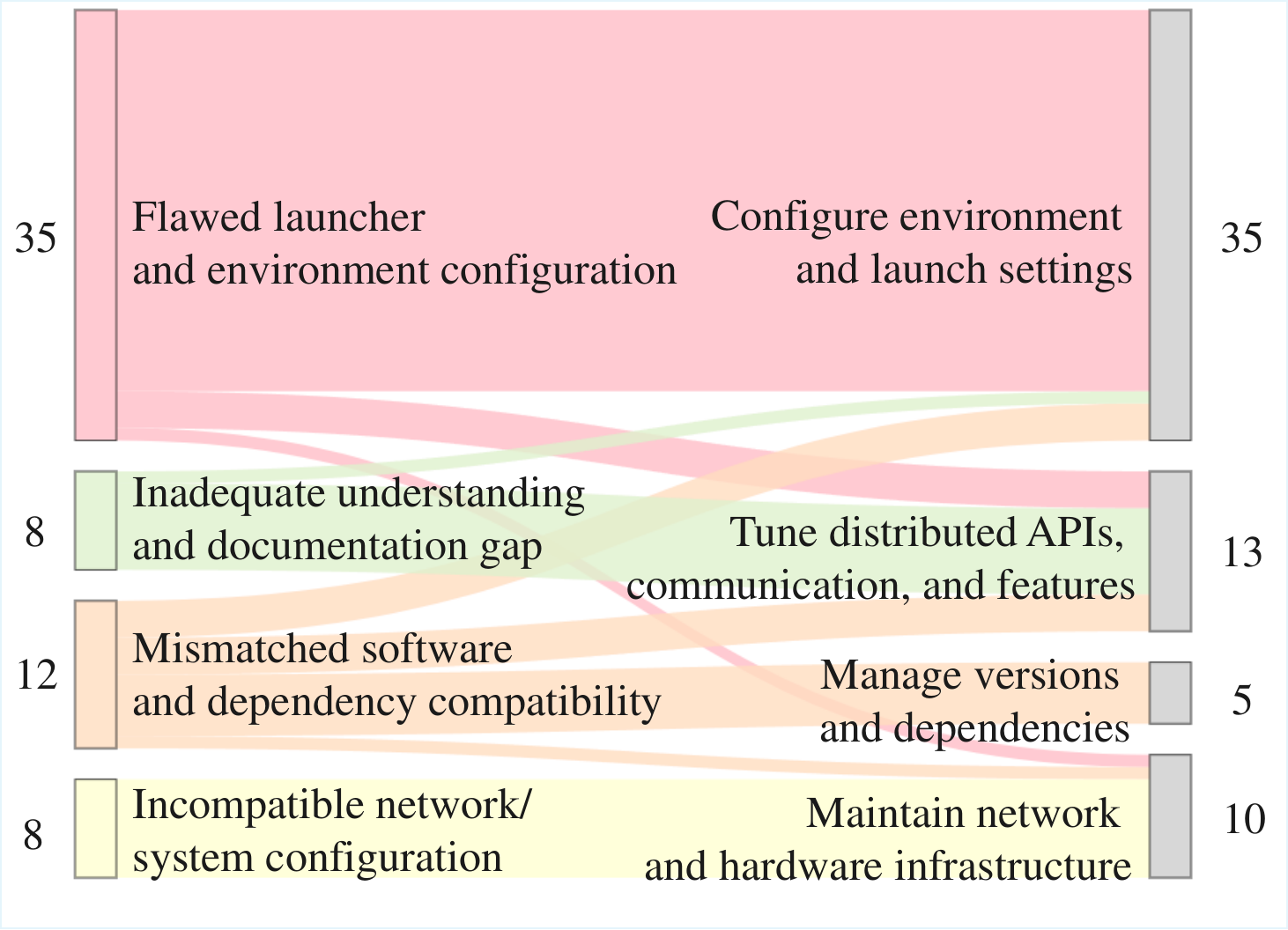}
    \hfill
    \includegraphics[width=0.32\linewidth]{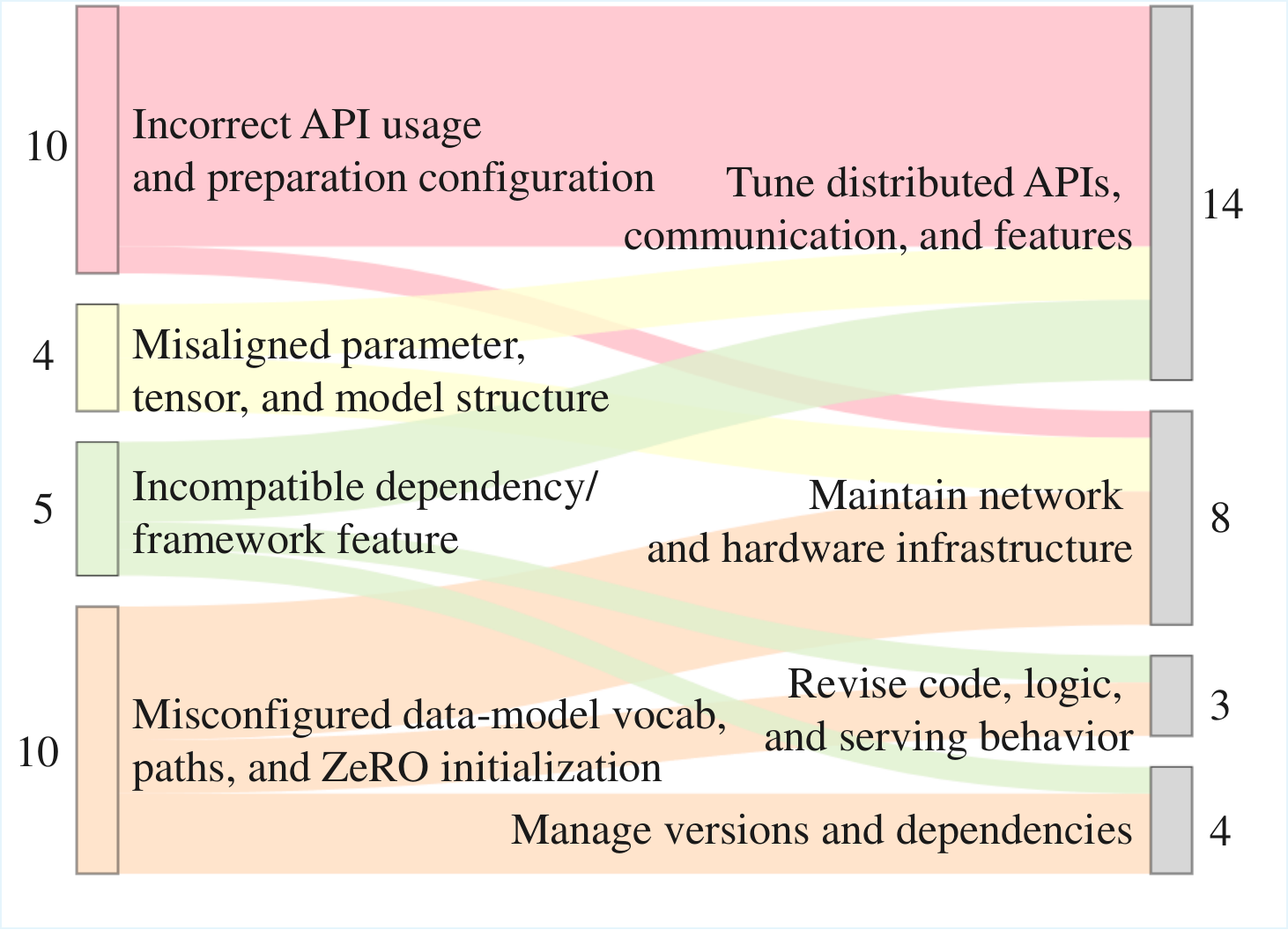}
    
    \includegraphics[width=0.48\linewidth]{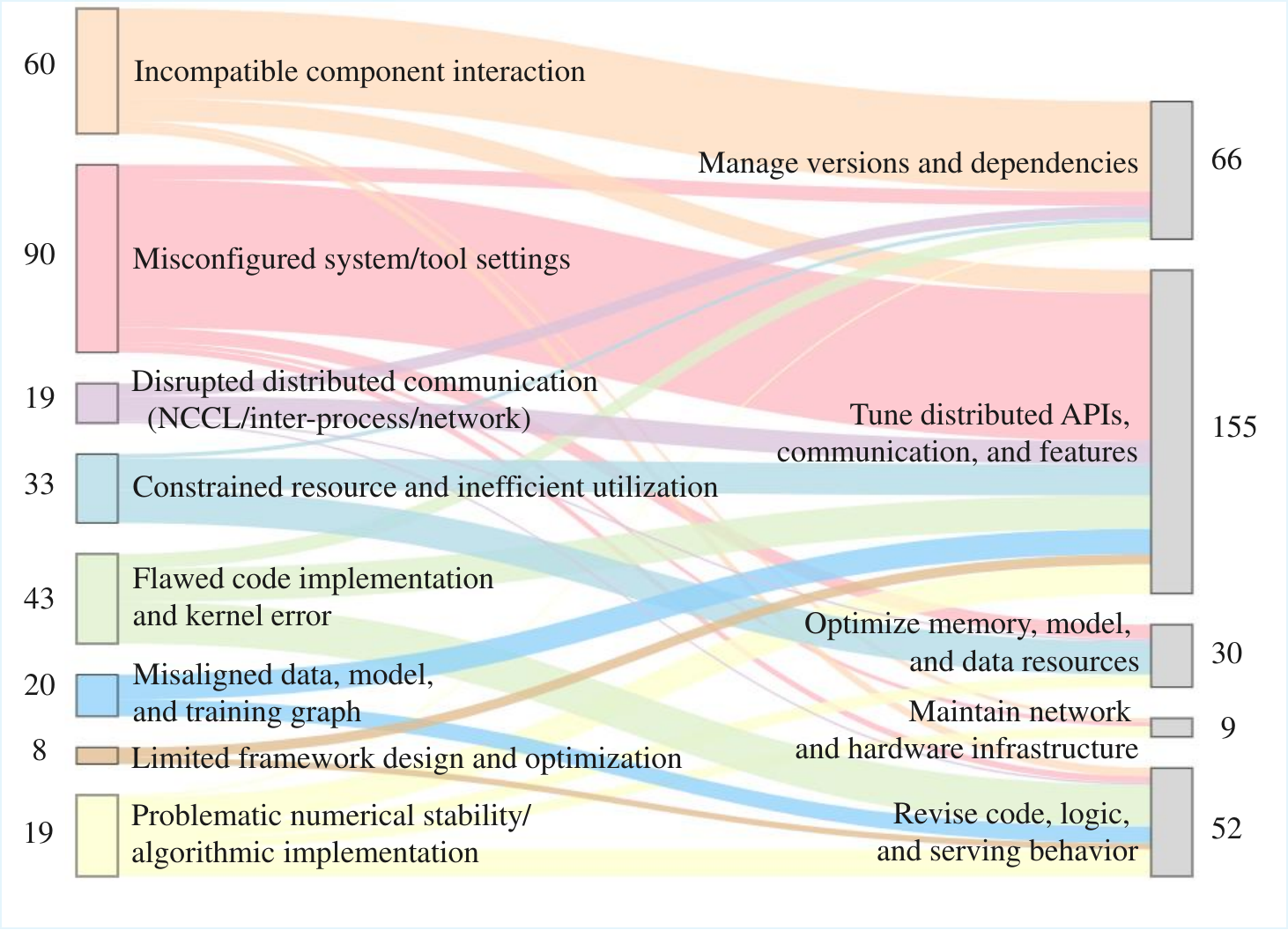}
    \hfill
    \includegraphics[width=0.48\linewidth]{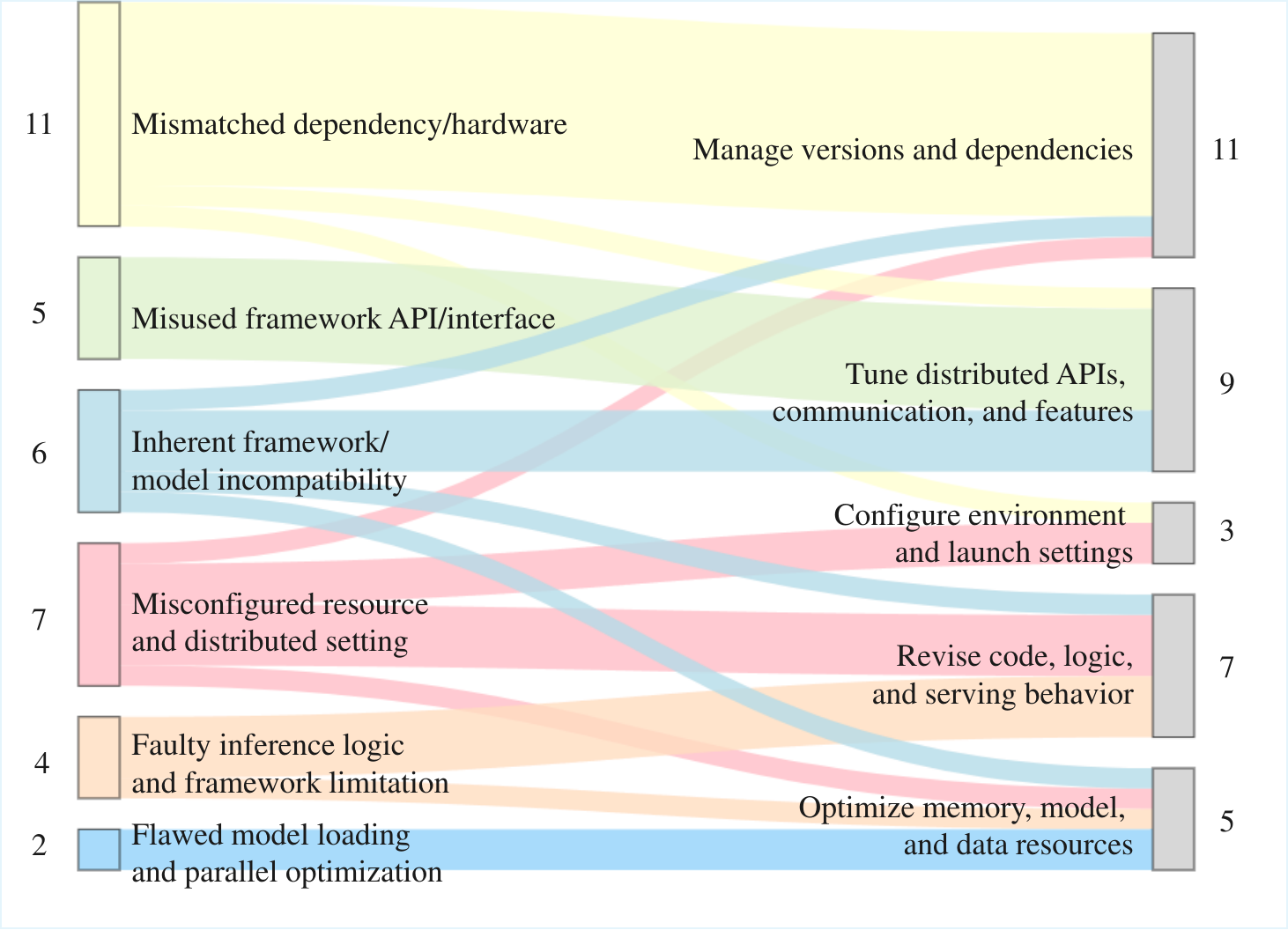}
\caption{Relationship between root causes and fix patterns.}
\label{fig:rc-fp}
\end{figure}

As demonstrated in Figure~\ref{fig:rc-fp}, all root causes of issues can be resolved using six common fix patterns: configure environment and launch settings (10.6\%); manage versions and dependencies (37.2\%), tuning distributed APIs, communication, and features (31.9\%), maintain network and hardware infrastructure (4.4\%), optimizing memory, model, and data resources (5.7\%), and revise code, logic, and serving behavior (10.1\%). 
Building on previous work, we further analyze the 25\% of issues that remain open for more than 67 days and receive over six comments, reflecting considerable debugging effort. For these particularly challenging cases, managing versions and dependencies, as well as tuning distributed APIs, communication, and features, are the two most effective fix patterns, successfully resolving 61.4\% of such issues. 

\begin{figure}[!th]
    \centering
    \includegraphics[width=0.7\linewidth]{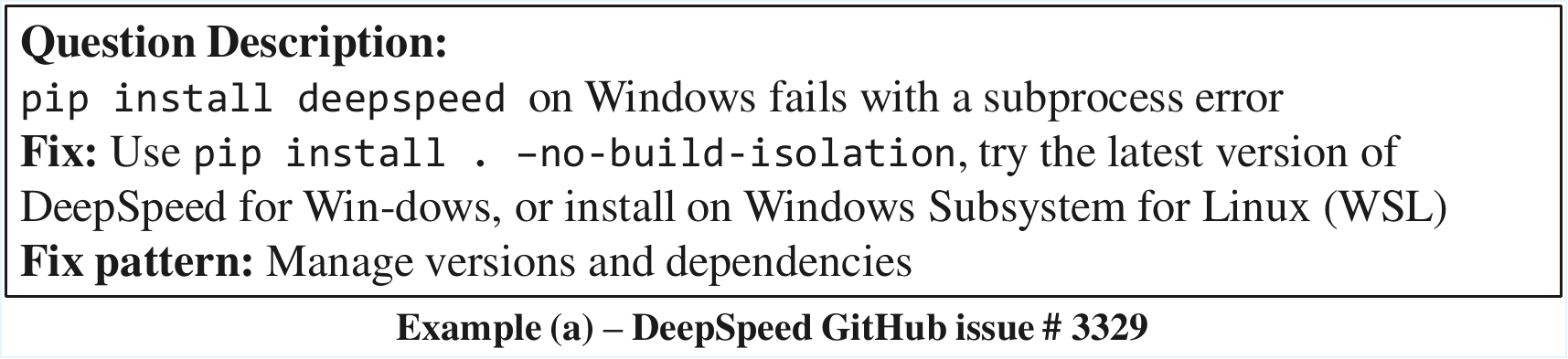}
    \hfill
    \includegraphics[width=0.7\linewidth]{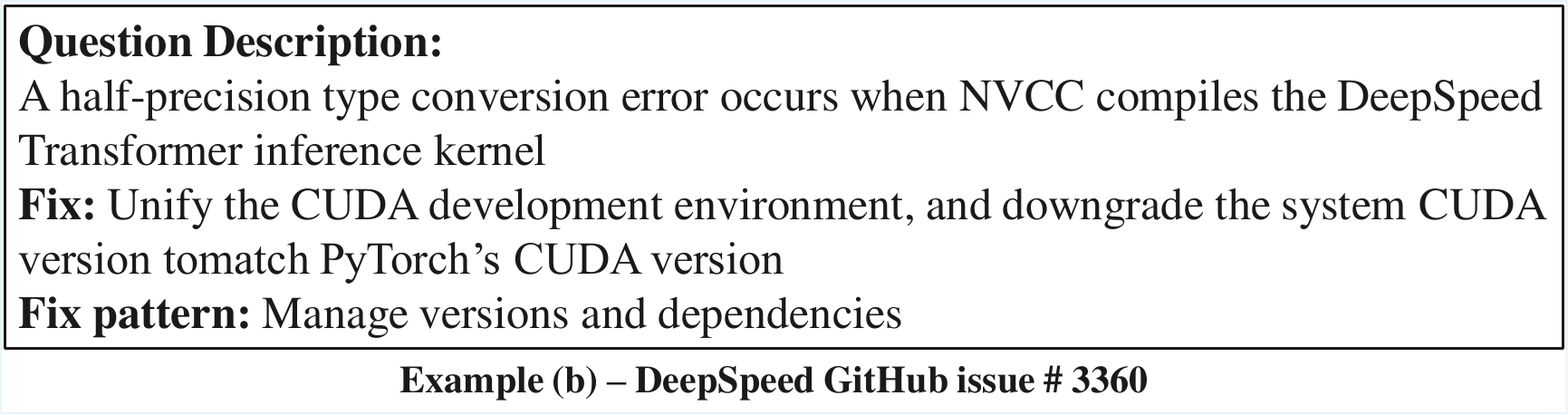}
    \hfill
    \includegraphics[width=0.7\linewidth]{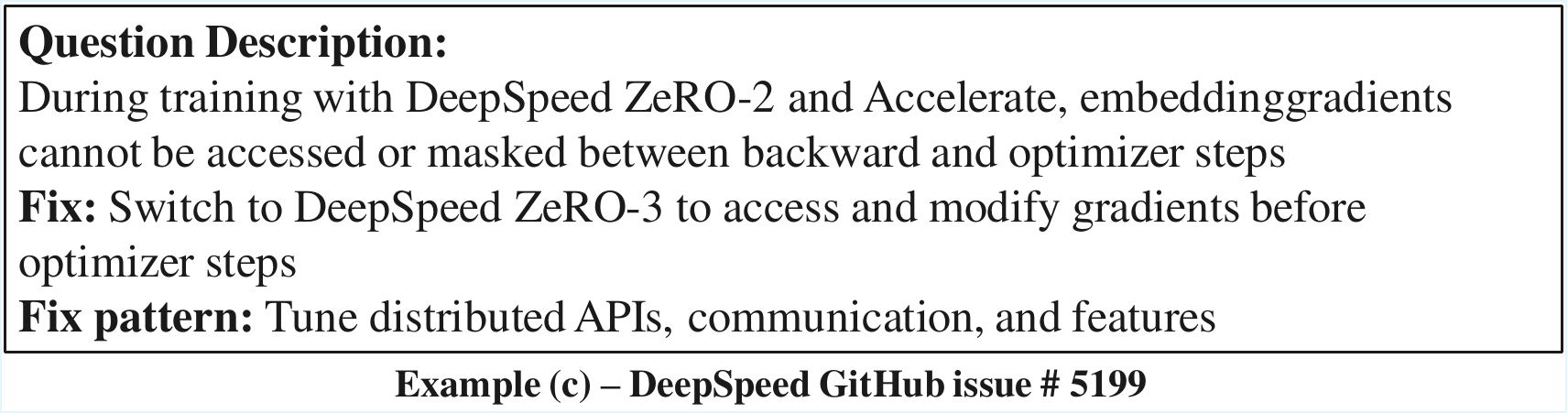}
    \hfill
    \includegraphics[width=0.7\linewidth]{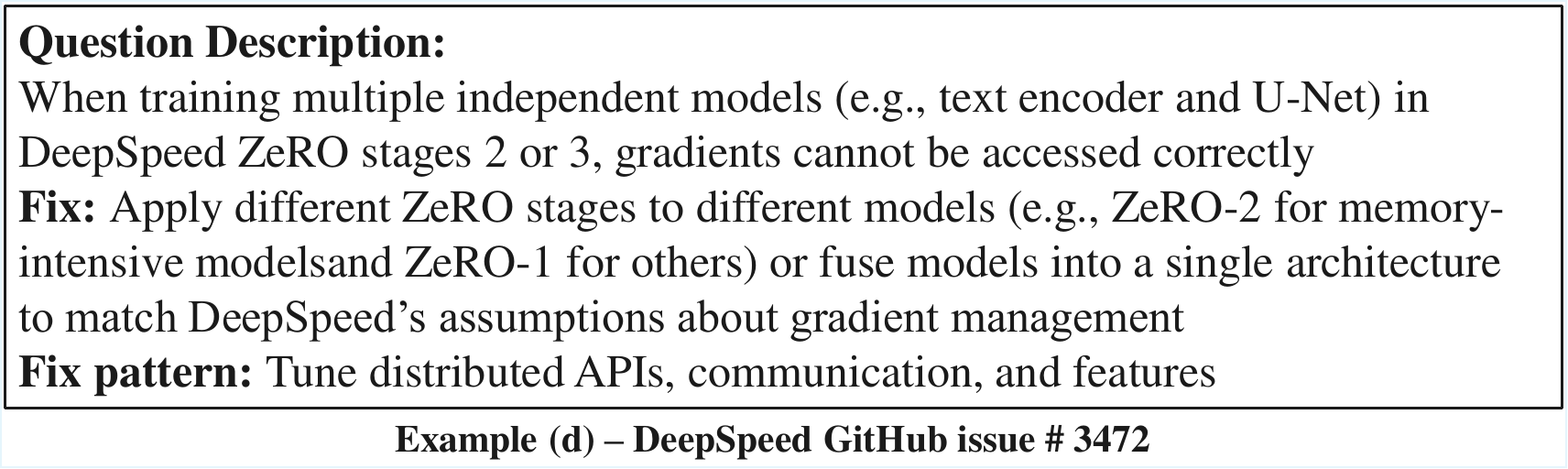}
    \caption{Examples of issues and their corresponding fix patterns.}
    \label{fig:examples for fp}
\end{figure}

In Figure~\ref{fig:examples for fp}, we present four examples, each having over 15 comments, to intuitively illustrate fix patterns. In Example (a)\cite{deepspeed_issue3329}, a user experienced an \textit{installation \& build failure (1.1)} when trying to install DeepSpeed on Windows. It was due to an inadequate build environment, such as misconfigured CMake, NVCC, or GCC/linker settings.
On Windows, pip's isolated builds could prevent PyTorch from being detected, and missing Linux-only dependencies like libaio could cause compilation failures. The main solution is to update the framework version or use the framework on WSL, because DeepSpeed is designed for Linux, so native Windows installation is challenging and poorly support. As a result, Windows installations lack comprehensive official testing and guidance, placing the majority of the troubleshooting burden on users. 
Resolving these issues often requires advanced knowledge of the operating system, compilers, and the inner workings of DeepSpeed. In some instances, users may even need to manually adjust DeepSpeed’s C++ source code to resolve Windows-specific compiler problems, which poses a very high bar for most practitioners. 
Similarly, Example (b)\cite{deepspeed_issue3360} reports an \textit{installation \& build failure (1.1)} during the compilation of DeepSpeed’s CUDA extensions.
The CUDA version used by PyTorch (11.7) differed from the system's CUDA version (12.1), which caused the CUDA compiler to fail on half-precision type conversions while compiling DeepSpeed’s kernel. DeepSpeed requires exact CUDA version compatibility for native extensions, stricter than PyTorch's minor version flexibility. Users, often unaware of this, may spend considerable time on trial-and-error fixes like changing macro definitions in \texttt{cpp\_extension.py} or build settings. Resolving this issue requires accurate environment matching and sometimes advanced modifications, making precise version control critical for reliable compilation.

Example (c)\cite{deepspeed_issue5199} presents a \textit{runtime error (4.3)} in DeepSpeed ZeRO-2 when used with Hugging Face Accelerate, where gradients cannot be accessed between the backward and optimizer steps. This limitation arises from code implementation flaws and kernel-level issues related to in-place operations during distributed communication. Because DeepSpeed is accessed indirectly through Accelerate's wrapper around \texttt{transformers.Trainer}, the underlying gradient workflow is abstracted away, making it difficult for users to control or observe these operations. Switching to ZeRO-3 can resolve this issue. However, the absence of explicit APIs and detailed documentation from DeepSpeed results in inefficient problem resolution for most users.
Example (d)\cite{deepspeed_issue3472} also reports a \textit{runtime error (4.3)} when training multiple models.  
This challenge arises because DeepSpeed's ZeRO optimizer (particularly stages 2 and 3) is primarily designed to optimize memory usage for a single, large model. Its implementation of gradient partitioning and memory management assumes that all models are uniformly managed by a single DeepSpeed engine, which breaks down when each model runs under a separate engine. Alternatives proposed by contributors (e.g., using stage 2 for the memory-intensive model and stage 1 for another model, or fusing model architectures) are not general solutions within the DeepSpeed framework. This indicates that DeepSpeed's current design inherently struggles to support direct multi-model training with ZeRO-3.

\textbf{For RQ3, see Findings F.7 and I.7 in Table~\ref{table:1}.}





\section{Threats to Validity}
We acknowledge several threats to validity in our study and have taken steps to mitigate them. To address data source selection bias, we initially chose Stack Overflow and GitHub, the two most widely used platforms in software engineering \cite{aghajani2019software}. This strategy is used to maximize the representativeness of collected distributed deep learning issues and capture a broad spectrum of practitioner challenges. However, during data collection, we observed substantially fewer distributed framework issues on Stack Overflow compared to GitHub. Accordingly, to maintain both rigor and broad coverage, we focused exclusively on GitHub issues for final analysis, ensuring a robust and practitioner-relevant dataset. 
To minimize bias in framework selection and issue relevance, we concentrated on three widely adopted open-source distributed deep learning frameworks and applied multi-stage filtering: formal labels were used to exclude irrelevant issues, followed by manual screening to retain only substantive bug-related issues. 
For manual annotation, we reduced subjectivity by having two authors independently code each issue using open coding, with all disagreements resolved by a third author. We validated our taxonomy through consensus and measured inter-rater reliability with Cohen’s Kappa, which showed very strong agreement (e.g., a high kappa value). Finally, recognizing that objectively assessing the difficulty of fix patterns is challenging, we used issue open duration and comment count as practical proxies for resolution complexity. While these metrics could be affected by community engagement, they offer practical indicators of the challenges practitioners face.

\section{Related Work}
\textbf{Distributed deep learning.}
Deep learning is quickly becoming popular in various fields, such as computer vision and speech recognition. Training deep learning models generally demands significant computational resources and relies on high-performance, costly GPUs. To manage the increasing volume of data, distributed deep learning (DDL) training is often utilized to harness the power of multiple GPUs in parallel \cite{gu2019tiresias}. 
Data parallelism \cite{krizhevsky2012imagenet, li2014communication, rajbhandari2020zero, sergeev2018horovod} is the most widely adopted strategy for training in DDL frameworks. Each worker trains a complete copy of the model on a specific portion of the dataset, and gradients are synchronized across all workers to achieve consistent model updates. This approach accelerates convergence, especially in synchronous mode, and enables efficient scaling with increasing numbers of GPUs. 
Model parallelism \cite{dean2012large} divides the model itself among multiple devices, which is essential when working with networks that cannot fit into the memory of a single GPU. Careful coordination between devices is required to maintain accurate parameter updates. Pipeline parallelism \cite{huang2019gpipe} further helps with memory limitations by separating models into sequential stages and processing mini-batches as staggered micro-batches. This allows activations to be passed to the next stage quickly and improves resource utilization.
In practice, hybrid parallelism \cite{fan2021dapple, jia2022whale,park2020hetpipe} (seen in frameworks like DeepSpeed \cite{rasley2020deepspeed}, Megatron-LM \cite{narayanan2021efficient}) combines pipeline and data parallelism,
improving throughput, resource efficiency, and scalability when training deep learning models.
\\
\textbf{Empirical study on bugs.}
Studies on framework bugs \cite{chen2023toward, li2023understanding, ho2023empirical, tambon2024silent,long2022reporting, islam2020repairing, jia2020empirical, jia2021symptoms, yang2022comprehensive} analyze issues within DL frameworks themselves. For example, Chen et al. \cite{chen2023toward} examined 1,000 bugs across four frameworks (e.g., TensorFlow, PyTorch) to identify common symptoms, root causes, and vulnerable layers. Islam et al. \cite{islam2020repairing} studied repairs from GitHub and Stack Overflow, investigating challenges and recurring repair patterns in popular deep learning libraries.
On the other hand, studies on framework usage bugs \cite{zhang2018empirical, islam2019comprehensive, chen2020comprehensive, humbatova2020taxonomy, zhang2020empirical, cao2022understanding, wang2023compatibility, liu2023rise} examine issues encountered when developers use deep learning frameworks. For instance, Zhang et al. \cite{zhang2018empirical} collected 175 TensorFlow-related bugs to analyze detection challenges, but were limited by framework scope and dataset size. Islam et al. \cite{islam2019comprehensive} covered multiple frameworks and identified bug-prone deep learning stages, but did not connect bugs to root causes or fix strategies. 
Chen et al. \cite{chen2020comprehensive} categorized 72 types of deployment bugs from Stack Overflow, offering practical insights for improving deep learning software but focused mainly on bug classification. Humbatova et al. \cite{humbatova2020taxonomy} analyzed 1059 real-world faults from GitHub and Stack Overflow for top deep learning frameworks and conducted interviews, expanding fault taxonomy but without deeply investigating causes or solutions.
Liu et al. \cite{liu2023rise} examined questions, symptoms, and fix strategies for distributed training faults across three general-purpose and one distributed deep learning frameworks, but did not differentiate which bugs are unique to each framework type or explain underlying causes. In contrast, our study addresses these gaps in two key ways: First, we categorize bugs in distributed frameworks and statistically determine whether each bug is unique to distributed systems or also present in general-purpose deep learning frameworks. Second, we conduct a comprehensive, stage-aware mapping that links symptoms, root causes, and fix patterns to each distinct stage of the distributed deep learning pipeline. Our study provides a more structured and actionable understanding of bug characteristics tailored to the unique workflow of distributed deep learning systems.

\section{Conclusion}
This study provides a comprehensive analysis of bug-related issues in distributed deep learning frameworks, based on manually identified cases from GitHub. We developed detailed taxonomies covering 34 core symptoms and 28 root causes, mapped specifically to stages of distributed training and inference. Our quantitative analysis highlights common fix patterns and presents examples of complex bugs. They are characterized by longer open durations and more discussion, which often pose greater challenges for resolution. These findings yield practical insights for framework developers and users, helping to enhance the robustness and reliability of both developing and applying distributed deep learning systems.

The data that support the findings of this study are available form the following link:
\url{https://figshare.com/articles/dataset/Bugs\_in\_Modern\_Distributed\_Deep\_Learning\_Systems/30104959?file=57874069}


\bibliographystyle{ACM-Reference-Format}

\bibliography{sample-base}


\begin{thebibliography}{94}


\ifx \showCODEN    \undefined \def \showCODEN     #1{\unskip}     \fi
\ifx \showISBNx    \undefined \def \showISBNx     #1{\unskip}     \fi
\ifx \showISBNxiii \undefined \def \showISBNxiii  #1{\unskip}     \fi
\ifx \showISSN     \undefined \def \showISSN      #1{\unskip}     \fi
\ifx \showLCCN     \undefined \def \showLCCN      #1{\unskip}     \fi
\ifx \shownote     \undefined \def \shownote      #1{#1}          \fi
\ifx \showarticletitle \undefined \def \showarticletitle #1{#1}   \fi
\ifx \showURL      \undefined \def \showURL       {\relax}        \fi
\providecommand\bibfield[2]{#2}
\providecommand\bibinfo[2]{#2}
\providecommand\natexlab[1]{#1}
\providecommand\showeprint[2][]{arXiv:#2}

\bibitem[Bug(2025)]%
        {Bugissue}
 \bibinfo{year}{2025}\natexlab{}.
\newblock \bibinfo{title}{Bug issue dataset}.
\newblock \bibinfo{howpublished}{Available: \url{https://figshare.com/articles/dataset/Bugs_in_Modern_Distributed_Deep_Learning_Systems/30104959?file=57874069}}.
\newblock
\newblock
\shownote{Accessed: [2025.9.9]. [Online]}.


\bibitem[Abadi et~al\mbox{.}(2016)]%
        {abadi2016tensorflow}
\bibfield{author}{\bibinfo{person}{Mart{\'\i}n Abadi}, \bibinfo{person}{Paul Barham}, \bibinfo{person}{Jianmin Chen}, \bibinfo{person}{Zhifeng Chen}, \bibinfo{person}{Andy Davis}, \bibinfo{person}{Jeffrey Dean}, \bibinfo{person}{Matthieu Devin}, \bibinfo{person}{Sanjay Ghemawat}, \bibinfo{person}{Geoffrey Irving}, \bibinfo{person}{Michael Isard}, {et~al\mbox{.}}} \bibinfo{year}{2016}\natexlab{}.
\newblock \showarticletitle{$\{$TensorFlow$\}$: a system for $\{$Large-Scale$\}$ machine learning}. In \bibinfo{booktitle}{\emph{12th USENIX symposium on operating systems design and implementation (OSDI 16)}}. \bibinfo{pages}{265--283}.
\newblock


\bibitem[Aghajani et~al\mbox{.}(2019)]%
        {aghajani2019software}
\bibfield{author}{\bibinfo{person}{Emad Aghajani}, \bibinfo{person}{Csaba Nagy}, \bibinfo{person}{Olga~Lucero Vega-M{\'a}rquez}, \bibinfo{person}{Mario Linares-V{\'a}squez}, \bibinfo{person}{Laura Moreno}, \bibinfo{person}{Gabriele Bavota}, {and} \bibinfo{person}{Michele Lanza}.} \bibinfo{year}{2019}\natexlab{}.
\newblock \showarticletitle{Software documentation issues unveiled}. In \bibinfo{booktitle}{\emph{2019 IEEE/ACM 41st International Conference on Software Engineering (ICSE)}}. IEEE, \bibinfo{pages}{1199--1210}.
\newblock


\bibitem[airlsyn(2024)]%
        {colossalAI_issue5290}
\bibfield{author}{\bibinfo{person}{airlsyn}.} \bibinfo{year}{2024}\natexlab{}.
\newblock \bibinfo{title}{Issue 5290 in colossalAI}.
\newblock \bibinfo{howpublished}{Available: \url{https://github.com/hpcaitech/ColossalAI/issues/5290}}.
\newblock
\newblock
\shownote{Accessed: [2025.9.9]. [Online]}.


\bibitem[Aminabadi et~al\mbox{.}(2022)]%
        {aminabadi2022deepspeed}
\bibfield{author}{\bibinfo{person}{Reza~Yazdani Aminabadi}, \bibinfo{person}{Samyam Rajbhandari}, \bibinfo{person}{Ammar~Ahmad Awan}, \bibinfo{person}{Cheng Li}, \bibinfo{person}{Du Li}, \bibinfo{person}{Elton Zheng}, \bibinfo{person}{Olatunji Ruwase}, \bibinfo{person}{Shaden Smith}, \bibinfo{person}{Minjia Zhang}, \bibinfo{person}{Jeff Rasley}, {et~al\mbox{.}}} \bibinfo{year}{2022}\natexlab{}.
\newblock \showarticletitle{Deepspeed-inference: enabling efficient inference of transformer models at unprecedented scale}. In \bibinfo{booktitle}{\emph{SC22: International Conference for High Performance Computing, Networking, Storage and Analysis}}. IEEE, \bibinfo{pages}{1--15}.
\newblock


\bibitem[AngThanos(2021)]%
        {deepspeed_issue1460}
\bibfield{author}{\bibinfo{person}{AngThanos}.} \bibinfo{year}{2021}\natexlab{}.
\newblock \bibinfo{title}{Issue 1460 in DeepSpeed}.
\newblock \bibinfo{howpublished}{Available: \url{https://github.com/deepspeedai/DeepSpeed/issues/1466}}.
\newblock
\newblock
\shownote{Accessed: [2025.9.9]. [Online]}.


\bibitem[Bachstelze(2022)]%
        {deepspeed_issue2558}
\bibfield{author}{\bibinfo{person}{Bachstelze}.} \bibinfo{year}{2022}\natexlab{}.
\newblock \bibinfo{title}{Issue 2558 in DeepSpeed}.
\newblock \bibinfo{howpublished}{Available: \url{https://github.com/deepspeedai/DeepSpeed/issues/2558}}.
\newblock
\newblock
\shownote{Accessed: [2025.9.9]. [Online]}.


\bibitem[bmedishe(2022)]%
        {deepspeed_issue1816}
\bibfield{author}{\bibinfo{person}{bmedishe}.} \bibinfo{year}{2022}\natexlab{}.
\newblock \bibinfo{title}{Issue 1816 in DeepSpeed}.
\newblock \bibinfo{howpublished}{Available: \url{https://github.com/deepspeedai/DeepSpeed/issues/1816}}.
\newblock
\newblock
\shownote{Accessed: [2025.9.9]. [Online]}.


\bibitem[Borzunov et~al\mbox{.}(2023)]%
        {borzunov2023distributed}
\bibfield{author}{\bibinfo{person}{Alexander Borzunov}, \bibinfo{person}{Max Ryabinin}, \bibinfo{person}{Artem Chumachenko}, \bibinfo{person}{Dmitry Baranchuk}, \bibinfo{person}{Tim Dettmers}, \bibinfo{person}{Younes Belkada}, \bibinfo{person}{Pavel Samygin}, {and} \bibinfo{person}{Colin~A Raffel}.} \bibinfo{year}{2023}\natexlab{}.
\newblock \showarticletitle{Distributed inference and fine-tuning of large language models over the internet}.
\newblock \bibinfo{journal}{\emph{Advances in neural information processing systems}}  \bibinfo{volume}{36} (\bibinfo{year}{2023}), \bibinfo{pages}{12312--12331}.
\newblock


\bibitem[Cao et~al\mbox{.}(2022)]%
        {cao2022understanding}
\bibfield{author}{\bibinfo{person}{Junming Cao}, \bibinfo{person}{Bihuan Chen}, \bibinfo{person}{Chao Sun}, \bibinfo{person}{Longjie Hu}, \bibinfo{person}{Shuaihong Wu}, {and} \bibinfo{person}{Xin Peng}.} \bibinfo{year}{2022}\natexlab{}.
\newblock \showarticletitle{Understanding performance problems in deep learning systems}. In \bibinfo{booktitle}{\emph{Proceedings of the 30th ACM Joint European Software Engineering Conference and Symposium on the Foundations of Software Engineering}}. \bibinfo{pages}{357--369}.
\newblock


\bibitem[Chen et~al\mbox{.}(2023)]%
        {chen2023toward}
\bibfield{author}{\bibinfo{person}{Junjie Chen}, \bibinfo{person}{Yihua Liang}, \bibinfo{person}{Qingchao Shen}, \bibinfo{person}{Jiajun Jiang}, {and} \bibinfo{person}{Shuochuan Li}.} \bibinfo{year}{2023}\natexlab{}.
\newblock \showarticletitle{Toward understanding deep learning framework bugs}.
\newblock \bibinfo{journal}{\emph{ACM Transactions on Software Engineering and Methodology}} \bibinfo{volume}{32}, \bibinfo{number}{6} (\bibinfo{year}{2023}), \bibinfo{pages}{1--31}.
\newblock


\bibitem[Chen et~al\mbox{.}(2020)]%
        {chen2020comprehensive}
\bibfield{author}{\bibinfo{person}{Zhenpeng Chen}, \bibinfo{person}{Yanbin Cao}, \bibinfo{person}{Yuanqiang Liu}, \bibinfo{person}{Haoyu Wang}, \bibinfo{person}{Tao Xie}, {and} \bibinfo{person}{Xuanzhe Liu}.} \bibinfo{year}{2020}\natexlab{}.
\newblock \showarticletitle{A comprehensive study on challenges in deploying deep learning based software}. In \bibinfo{booktitle}{\emph{Proceedings of the 28th ACM joint meeting on European software engineering conference and symposium on the foundations of software engineering}}. \bibinfo{pages}{750--762}.
\newblock


\bibitem[chunyang wen(2021)]%
        {deepspeed_issue1523}
\bibfield{author}{\bibinfo{person}{chunyang wen}.} \bibinfo{year}{2021}\natexlab{}.
\newblock \bibinfo{title}{Issue 1523 in DeepSpeed}.
\newblock \bibinfo{howpublished}{Available: \url{https://github.com/deepspeedai/DeepSpeed/issues/1523}}.
\newblock
\newblock
\shownote{Accessed: [2025.9.9]. [Online]}.


\bibitem[crazycth(2023)]%
        {deepspeed_issue2895}
\bibfield{author}{\bibinfo{person}{crazycth}.} \bibinfo{year}{2023}\natexlab{}.
\newblock \bibinfo{title}{Issue 2895 in DeepSpeed}.
\newblock \bibinfo{howpublished}{Available: \url{https://github.com/deepspeedai/DeepSpeed/issues/2895}}.
\newblock
\newblock
\shownote{Accessed: [2025.9.9]. [Online]}.


\bibitem[Dai et~al\mbox{.}(2022)]%
        {dai2022reveal}
\bibfield{author}{\bibinfo{person}{Hulin Dai}, \bibinfo{person}{Xuan Peng}, \bibinfo{person}{Xuanhua Shi}, \bibinfo{person}{Ligang He}, \bibinfo{person}{Qian Xiong}, {and} \bibinfo{person}{Hai Jin}.} \bibinfo{year}{2022}\natexlab{}.
\newblock \showarticletitle{Reveal training performance mystery between TensorFlow and PyTorch in the single GPU environment}.
\newblock \bibinfo{journal}{\emph{Science China Information Sciences}} \bibinfo{volume}{65}, \bibinfo{number}{1} (\bibinfo{year}{2022}), \bibinfo{pages}{112103}.
\newblock


\bibitem[Dean et~al\mbox{.}(2012)]%
        {dean2012large}
\bibfield{author}{\bibinfo{person}{Jeffrey Dean}, \bibinfo{person}{Greg Corrado}, \bibinfo{person}{Rajat Monga}, \bibinfo{person}{Kai Chen}, \bibinfo{person}{Matthieu Devin}, \bibinfo{person}{Mark Mao}, \bibinfo{person}{Marc'aurelio Ranzato}, \bibinfo{person}{Andrew Senior}, \bibinfo{person}{Paul Tucker}, \bibinfo{person}{Ke Yang}, {et~al\mbox{.}}} \bibinfo{year}{2012}\natexlab{}.
\newblock \showarticletitle{Large scale distributed deep networks}.
\newblock \bibinfo{journal}{\emph{Advances in neural information processing systems}}  \bibinfo{volume}{25} (\bibinfo{year}{2012}).
\newblock


\bibitem[delock(2021)]%
        {deepspeed_issue1364}
\bibfield{author}{\bibinfo{person}{delock}.} \bibinfo{year}{2021}\natexlab{}.
\newblock \bibinfo{title}{Issue 1364 in DeepSpeed}.
\newblock \bibinfo{howpublished}{Available: \url{https://github.com/deepspeedai/DeepSpeed/issues/1364}}.
\newblock
\newblock
\shownote{Accessed: [2025.9.9]. [Online]}.


\bibitem[dianyo(2022)]%
        {deepspeed_issue2608}
\bibfield{author}{\bibinfo{person}{dianyo}.} \bibinfo{year}{2022}\natexlab{}.
\newblock \bibinfo{title}{Issue 2608 in DeepSpeed}.
\newblock \bibinfo{howpublished}{Available: \url{https://github.com/deepspeedai/DeepSpeed/issues/2608}}.
\newblock
\newblock
\shownote{Accessed: [2025.9.9]. [Online]}.


\bibitem[dogacancolak(2022)]%
        {deepspeed_issue2632}
\bibfield{author}{\bibinfo{person}{dogacancolak}.} \bibinfo{year}{2022}\natexlab{}.
\newblock \bibinfo{title}{Issue 2632 in DeepSpeed}.
\newblock \bibinfo{howpublished}{Available: \url{https://github.com/deepspeedai/DeepSpeed/issues/2632}}.
\newblock
\newblock
\shownote{Accessed: [2025.9.9]. [Online]}.


\bibitem[Fan et~al\mbox{.}(2021)]%
        {fan2021dapple}
\bibfield{author}{\bibinfo{person}{Shiqing Fan}, \bibinfo{person}{Yi Rong}, \bibinfo{person}{Chen Meng}, \bibinfo{person}{Zongyan Cao}, \bibinfo{person}{Siyu Wang}, \bibinfo{person}{Zhen Zheng}, \bibinfo{person}{Chuan Wu}, \bibinfo{person}{Guoping Long}, \bibinfo{person}{Jun Yang}, \bibinfo{person}{Lixue Xia}, {et~al\mbox{.}}} \bibinfo{year}{2021}\natexlab{}.
\newblock \showarticletitle{DAPPLE: A pipelined data parallel approach for training large models}. In \bibinfo{booktitle}{\emph{Proceedings of the 26th ACM SIGPLAN Symposium on Principles and Practice of Parallel Programming}}. \bibinfo{pages}{431--445}.
\newblock


\bibitem[feiliya333(2023)]%
        {deepspeed_issue3235}
\bibfield{author}{\bibinfo{person}{feiliya333}.} \bibinfo{year}{2023}\natexlab{}.
\newblock \bibinfo{title}{Issue 3235 in DeepSpeed}.
\newblock \bibinfo{howpublished}{Available: \url{https://github.com/deepspeedai/DeepSpeed/issues/3235}}.
\newblock
\newblock
\shownote{Accessed: [2025.9.9]. [Online]}.


\bibitem[Gao et~al\mbox{.}(2020)]%
        {gao2020estimating}
\bibfield{author}{\bibinfo{person}{Yanjie Gao}, \bibinfo{person}{Yu Liu}, \bibinfo{person}{Hongyu Zhang}, \bibinfo{person}{Zhengxian Li}, \bibinfo{person}{Yonghao Zhu}, \bibinfo{person}{Haoxiang Lin}, {and} \bibinfo{person}{Mao Yang}.} \bibinfo{year}{2020}\natexlab{}.
\newblock \showarticletitle{Estimating GPU memory consumption of deep learning models}. In \bibinfo{booktitle}{\emph{Proceedings of the 28th ACM Joint Meeting on European Software Engineering Conference and Symposium on the Foundations of Software Engineering}}. \bibinfo{pages}{1342--1352}.
\newblock


\bibitem[gnovack(2023)]%
        {deepspeed_issue3734}
\bibfield{author}{\bibinfo{person}{gnovack}.} \bibinfo{year}{2023}\natexlab{}.
\newblock \bibinfo{title}{Issue 3734 in DeepSpeed}.
\newblock \bibinfo{howpublished}{Available: \url{https://github.com/deepspeedai/DeepSpeed/issues/3734}}.
\newblock
\newblock
\shownote{Accessed: [2025.9.9]. [Online]}.


\bibitem[gongbudaizhe(2021)]%
        {deepspeed_issue1356}
\bibfield{author}{\bibinfo{person}{gongbudaizhe}.} \bibinfo{year}{2021}\natexlab{}.
\newblock \bibinfo{title}{Issue 1356 in DeepSpeed}.
\newblock \bibinfo{howpublished}{Available: \url{https://github.com/deepspeedai/DeepSpeed/issues/1356}}.
\newblock
\newblock
\shownote{Accessed: [2025.9.9]. [Online]}.


\bibitem[groganz(2022)]%
        {colossalAI_issue1756}
\bibfield{author}{\bibinfo{person}{groganz}.} \bibinfo{year}{2022}\natexlab{}.
\newblock \bibinfo{title}{Issue 1756 in colossalAI}.
\newblock \bibinfo{howpublished}{Available: \url{https://github.com/hpcaitech/ColossalAI/issues/1756}}.
\newblock
\newblock
\shownote{Accessed: [2025.9.9]. [Online]}.


\bibitem[Gu et~al\mbox{.}(2019)]%
        {gu2019tiresias}
\bibfield{author}{\bibinfo{person}{Juncheng Gu}, \bibinfo{person}{Mosharaf Chowdhury}, \bibinfo{person}{Kang~G Shin}, \bibinfo{person}{Yibo Zhu}, \bibinfo{person}{Myeongjae Jeon}, \bibinfo{person}{Junjie Qian}, \bibinfo{person}{Hongqiang Liu}, {and} \bibinfo{person}{Chuanxiong Guo}.} \bibinfo{year}{2019}\natexlab{}.
\newblock \showarticletitle{Tiresias: A $\{$GPU$\}$ cluster manager for distributed deep learning}. In \bibinfo{booktitle}{\emph{16th USENIX Symposium on Networked Systems Design and Implementation (NSDI 19)}}. \bibinfo{pages}{485--500}.
\newblock


\bibitem[Gulli and Pal(2017)]%
        {gulli2017deep}
\bibfield{author}{\bibinfo{person}{Antonio Gulli} {and} \bibinfo{person}{Sujit Pal}.} \bibinfo{year}{2017}\natexlab{}.
\newblock \bibinfo{booktitle}{\emph{Deep learning with Keras}}.
\newblock \bibinfo{publisher}{Packt Publishing Ltd}.
\newblock


\bibitem[guohe369(2023)]%
        {colossalAI_issue3109}
\bibfield{author}{\bibinfo{person}{guohe369}.} \bibinfo{year}{2023}\natexlab{}.
\newblock \bibinfo{title}{Issue 3109 in colossalAI}.
\newblock \bibinfo{howpublished}{Available: \url{https://github.com/hpcaitech/ColossalAI/issues/3109}}.
\newblock
\newblock
\shownote{Accessed: [2025.9.9]. [Online]}.


\bibitem[Ho et~al\mbox{.}(2023)]%
        {ho2023empirical}
\bibfield{author}{\bibinfo{person}{Sharon Chee~Yin Ho}, \bibinfo{person}{Vahid Majdinasab}, \bibinfo{person}{Mohayeminul Islam}, \bibinfo{person}{Diego~Elias Costa}, \bibinfo{person}{Emad Shihab}, \bibinfo{person}{Foutse Khomh}, \bibinfo{person}{Sarah Nadi}, {and} \bibinfo{person}{Muhammad Raza}.} \bibinfo{year}{2023}\natexlab{}.
\newblock \showarticletitle{An empirical study on bugs inside pytorch: A replication study}. In \bibinfo{booktitle}{\emph{2023 IEEE International Conference on Software Maintenance and Evolution (ICSME)}}. IEEE, \bibinfo{pages}{220--231}.
\newblock


\bibitem[Hong et~al\mbox{.}(2024)]%
        {hong2024investigating}
\bibfield{author}{\bibinfo{person}{Shuo Hong}, \bibinfo{person}{Hailong Sun}, \bibinfo{person}{Xiang Gao}, {and} \bibinfo{person}{Shin~Hwei Tan}.} \bibinfo{year}{2024}\natexlab{}.
\newblock \showarticletitle{Investigating and detecting silent bugs in pytorch programs}. In \bibinfo{booktitle}{\emph{2024 IEEE International Conference on Software Analysis, Evolution and Reengineering (SANER)}}. IEEE, \bibinfo{pages}{272--283}.
\newblock


\bibitem[Huang et~al\mbox{.}(2019)]%
        {huang2019gpipe}
\bibfield{author}{\bibinfo{person}{Yanping Huang}, \bibinfo{person}{Youlong Cheng}, \bibinfo{person}{Ankur Bapna}, \bibinfo{person}{Orhan Firat}, \bibinfo{person}{Dehao Chen}, \bibinfo{person}{Mia Chen}, \bibinfo{person}{HyoukJoong Lee}, \bibinfo{person}{Jiquan Ngiam}, \bibinfo{person}{Quoc~V Le}, \bibinfo{person}{Yonghui Wu}, {et~al\mbox{.}}} \bibinfo{year}{2019}\natexlab{}.
\newblock \showarticletitle{Gpipe: Efficient training of giant neural networks using pipeline parallelism}.
\newblock \bibinfo{journal}{\emph{Advances in neural information processing systems}}  \bibinfo{volume}{32} (\bibinfo{year}{2019}).
\newblock


\bibitem[Humbatova et~al\mbox{.}(2020)]%
        {humbatova2020taxonomy}
\bibfield{author}{\bibinfo{person}{Nargiz Humbatova}, \bibinfo{person}{Gunel Jahangirova}, \bibinfo{person}{Gabriele Bavota}, \bibinfo{person}{Vincenzo Riccio}, \bibinfo{person}{Andrea Stocco}, {and} \bibinfo{person}{Paolo Tonella}.} \bibinfo{year}{2020}\natexlab{}.
\newblock \showarticletitle{Taxonomy of real faults in deep learning systems}. In \bibinfo{booktitle}{\emph{Proceedings of the ACM/IEEE 42nd international conference on software engineering}}. \bibinfo{pages}{1110--1121}.
\newblock


\bibitem[iamsimha(2022)]%
        {deepspeed_issue2223}
\bibfield{author}{\bibinfo{person}{iamsimha}.} \bibinfo{year}{2022}\natexlab{}.
\newblock \bibinfo{title}{Issue 2223 in DeepSpeed}.
\newblock \bibinfo{howpublished}{Available: \url{https://github.com/deepspeedai/DeepSpeed/issues/2223}}.
\newblock
\newblock
\shownote{Accessed: [2025.9.9]. [Online]}.


\bibitem[ianbstewart(2022)]%
        {deepspeed_issue2607}
\bibfield{author}{\bibinfo{person}{ianbstewart}.} \bibinfo{year}{2022}\natexlab{}.
\newblock \bibinfo{title}{Issue 2607 in DeepSpeed}.
\newblock \bibinfo{howpublished}{Available: \url{https://github.com/deepspeedai/DeepSpeed/issues/2607}}.
\newblock
\newblock
\shownote{Accessed: [2025.9.9]. [Online]}.


\bibitem[Islam et~al\mbox{.}(2019)]%
        {islam2019comprehensive}
\bibfield{author}{\bibinfo{person}{Md~Johirul Islam}, \bibinfo{person}{Giang Nguyen}, \bibinfo{person}{Rangeet Pan}, {and} \bibinfo{person}{Hridesh Rajan}.} \bibinfo{year}{2019}\natexlab{}.
\newblock \showarticletitle{A comprehensive study on deep learning bug characteristics}. In \bibinfo{booktitle}{\emph{Proceedings of the 2019 27th ACM joint meeting on european software engineering conference and symposium on the foundations of software engineering}}. \bibinfo{pages}{510--520}.
\newblock


\bibitem[Islam et~al\mbox{.}(2020)]%
        {islam2020repairing}
\bibfield{author}{\bibinfo{person}{Md~Johirul Islam}, \bibinfo{person}{Rangeet Pan}, \bibinfo{person}{Giang Nguyen}, {and} \bibinfo{person}{Hridesh Rajan}.} \bibinfo{year}{2020}\natexlab{}.
\newblock \showarticletitle{Repairing deep neural networks: Fix patterns and challenges}. In \bibinfo{booktitle}{\emph{Proceedings of the ACM/IEEE 42nd international conference on software engineering}}. \bibinfo{pages}{1135--1146}.
\newblock


\bibitem[JerryAllison(2023)]%
        {deepspeed_issue3329}
\bibfield{author}{\bibinfo{person}{JerryAllison}.} \bibinfo{year}{2023}\natexlab{}.
\newblock \bibinfo{title}{Issue 3329 in DeepSpeed}.
\newblock \bibinfo{howpublished}{Available: \url{https://github.com/deepspeedai/DeepSpeed/issues/3329}}.
\newblock
\newblock
\shownote{Accessed: [2025.9.9]. [Online]}.


\bibitem[Jia et~al\mbox{.}(2020)]%
        {jia2020empirical}
\bibfield{author}{\bibinfo{person}{Li Jia}, \bibinfo{person}{Hao Zhong}, \bibinfo{person}{Xiaoyin Wang}, \bibinfo{person}{Linpeng Huang}, {and} \bibinfo{person}{Xuansheng Lu}.} \bibinfo{year}{2020}\natexlab{}.
\newblock \showarticletitle{An empirical study on bugs inside tensorflow}. In \bibinfo{booktitle}{\emph{International Conference on Database Systems for Advanced Applications}}. Springer, \bibinfo{pages}{604--620}.
\newblock


\bibitem[Jia et~al\mbox{.}(2021)]%
        {jia2021symptoms}
\bibfield{author}{\bibinfo{person}{Li Jia}, \bibinfo{person}{Hao Zhong}, \bibinfo{person}{Xiaoyin Wang}, \bibinfo{person}{Linpeng Huang}, {and} \bibinfo{person}{Xuansheng Lu}.} \bibinfo{year}{2021}\natexlab{}.
\newblock \showarticletitle{The symptoms, causes, and repairs of bugs inside a deep learning library}.
\newblock \bibinfo{journal}{\emph{Journal of Systems and Software}}  \bibinfo{volume}{177} (\bibinfo{year}{2021}), \bibinfo{pages}{110935}.
\newblock


\bibitem[Jia et~al\mbox{.}(2022)]%
        {jia2022whale}
\bibfield{author}{\bibinfo{person}{Xianyan Jia}, \bibinfo{person}{Le Jiang}, \bibinfo{person}{Ang Wang}, \bibinfo{person}{Wencong Xiao}, \bibinfo{person}{Ziji Shi}, \bibinfo{person}{Jie Zhang}, \bibinfo{person}{Xinyuan Li}, \bibinfo{person}{Langshi Chen}, \bibinfo{person}{Yong Li}, \bibinfo{person}{Zhen Zheng}, {et~al\mbox{.}}} \bibinfo{year}{2022}\natexlab{}.
\newblock \showarticletitle{Whale: Efficient giant model training over heterogeneous $\{$GPUs$\}$}. In \bibinfo{booktitle}{\emph{2022 USENIX Annual Technical Conference (USENIX ATC 22)}}. \bibinfo{pages}{673--688}.
\newblock


\bibitem[joehoover(2022)]%
        {deepspeed_issue1770}
\bibfield{author}{\bibinfo{person}{joehoover}.} \bibinfo{year}{2022}\natexlab{}.
\newblock \bibinfo{title}{Issue 1770 in DeepSpeed}.
\newblock \bibinfo{howpublished}{Available: \url{https://github.com/deepspeedai/DeepSpeed/issues/1770}}.
\newblock
\newblock
\shownote{Accessed: [2025.9.9]. [Online]}.


\bibitem[kamalkraj(2021)]%
        {deepspeed_issue1362}
\bibfield{author}{\bibinfo{person}{kamalkraj}.} \bibinfo{year}{2021}\natexlab{}.
\newblock \bibinfo{title}{Issue 1362 in DeepSpeed}.
\newblock \bibinfo{howpublished}{Available: \url{https://github.com/deepspeedai/DeepSpeed/pull/1362}}.
\newblock
\newblock
\shownote{Accessed: [2025.9.9]. [Online]}.


\bibitem[khalil Hennara(2024)]%
        {deepspeed_issue5199}
\bibfield{author}{\bibinfo{person}{khalil Hennara}.} \bibinfo{year}{2024}\natexlab{}.
\newblock \bibinfo{title}{Issue 5199 in DeepSpeed}.
\newblock \bibinfo{howpublished}{Available: \url{https://github.com/deepspeedai/DeepSpeed/issues/5199}}.
\newblock
\newblock
\shownote{Accessed: [2025.9.9]. [Online]}.


\bibitem[koking0(2022)]%
        {colossalAI_issue1402}
\bibfield{author}{\bibinfo{person}{koking0}.} \bibinfo{year}{2022}\natexlab{}.
\newblock \bibinfo{title}{Issue 1402 in colossalAI}.
\newblock \bibinfo{howpublished}{Available: \url{https://github.com/hpcaitech/ColossalAI/issues/1402}}.
\newblock
\newblock
\shownote{Accessed: [2025.9.9]. [Online]}.


\bibitem[Krizhevsky et~al\mbox{.}(2012)]%
        {krizhevsky2012imagenet}
\bibfield{author}{\bibinfo{person}{Alex Krizhevsky}, \bibinfo{person}{Ilya Sutskever}, {and} \bibinfo{person}{Geoffrey~E Hinton}.} \bibinfo{year}{2012}\natexlab{}.
\newblock \showarticletitle{Imagenet classification with deep convolutional neural networks}.
\newblock \bibinfo{journal}{\emph{Advances in neural information processing systems}}  \bibinfo{volume}{25} (\bibinfo{year}{2012}).
\newblock


\bibitem[lambda7xx(2023)]%
        {deepspeed_issue2917}
\bibfield{author}{\bibinfo{person}{lambda7xx}.} \bibinfo{year}{2023}\natexlab{}.
\newblock \bibinfo{title}{Issue 2917 in DeepSpeed}.
\newblock \bibinfo{howpublished}{Available: \url{https://github.com/deepspeedai/DeepSpeed/issues/2917}}.
\newblock
\newblock
\shownote{Accessed: [2025.9.9]. [Online]}.


\bibitem[LeCun et~al\mbox{.}(2015)]%
        {lecun2015deep}
\bibfield{author}{\bibinfo{person}{Yann LeCun}, \bibinfo{person}{Yoshua Bengio}, {and} \bibinfo{person}{Geoffrey Hinton}.} \bibinfo{year}{2015}\natexlab{}.
\newblock \showarticletitle{Deep learning}.
\newblock \bibinfo{journal}{\emph{nature}} \bibinfo{volume}{521}, \bibinfo{number}{7553} (\bibinfo{year}{2015}), \bibinfo{pages}{436--444}.
\newblock


\bibitem[Li et~al\mbox{.}(2024)]%
        {li2024deepspeed}
\bibfield{author}{\bibinfo{person}{Conglong Li}, \bibinfo{person}{Zhewei Yao}, \bibinfo{person}{Xiaoxia Wu}, \bibinfo{person}{Minjia Zhang}, \bibinfo{person}{Connor Holmes}, \bibinfo{person}{Cheng Li}, {and} \bibinfo{person}{Yuxiong He}.} \bibinfo{year}{2024}\natexlab{}.
\newblock \showarticletitle{Deepspeed data efficiency: Improving deep learning model quality and training efficiency via efficient data sampling and routing}. In \bibinfo{booktitle}{\emph{Proceedings of the AAAI Conference on Artificial Intelligence}}, Vol.~\bibinfo{volume}{38}. \bibinfo{pages}{18490--18498}.
\newblock


\bibitem[Li et~al\mbox{.}(2014)]%
        {li2014communication}
\bibfield{author}{\bibinfo{person}{Mu Li}, \bibinfo{person}{David~G Andersen}, \bibinfo{person}{Alexander Smola}, {and} \bibinfo{person}{Kai Yu}.} \bibinfo{year}{2014}\natexlab{}.
\newblock \showarticletitle{Communication efficient distributed machine learning with the parameter server}.
\newblock \bibinfo{journal}{\emph{Advances in neural information processing systems}}  \bibinfo{volume}{27} (\bibinfo{year}{2014}).
\newblock


\bibitem[Li et~al\mbox{.}(2023a)]%
        {li2023colossal}
\bibfield{author}{\bibinfo{person}{Shenggui Li}, \bibinfo{person}{Hongxin Liu}, \bibinfo{person}{Zhengda Bian}, \bibinfo{person}{Jiarui Fang}, \bibinfo{person}{Haichen Huang}, \bibinfo{person}{Yuliang Liu}, \bibinfo{person}{Boxiang Wang}, {and} \bibinfo{person}{Yang You}.} \bibinfo{year}{2023}\natexlab{a}.
\newblock \showarticletitle{Colossal-ai: A unified deep learning system for large-scale parallel training}. In \bibinfo{booktitle}{\emph{Proceedings of the 52nd International Conference on Parallel Processing}}. \bibinfo{pages}{766--775}.
\newblock


\bibitem[Li et~al\mbox{.}(2023b)]%
        {li2023understanding}
\bibfield{author}{\bibinfo{person}{Zengyang Li}, \bibinfo{person}{Sicheng Wang}, \bibinfo{person}{Wenshuo Wang}, \bibinfo{person}{Peng Liang}, \bibinfo{person}{Ran Mo}, {and} \bibinfo{person}{Bing Li}.} \bibinfo{year}{2023}\natexlab{b}.
\newblock \showarticletitle{Understanding bugs in multi-language deep learning frameworks}. In \bibinfo{booktitle}{\emph{2023 IEEE/ACM 31st International Conference on Program Comprehension (ICPC)}}. IEEE, \bibinfo{pages}{328--338}.
\newblock


\bibitem[lijieyuan(2023)]%
        {colossalAI_issue4412}
\bibfield{author}{\bibinfo{person}{lijieyuan}.} \bibinfo{year}{2023}\natexlab{}.
\newblock \bibinfo{title}{Issue 4412 in colossalAI}.
\newblock \bibinfo{howpublished}{Available: \url{https://github.com/hpcaitech/ColossalAI/issues/4412}}.
\newblock
\newblock
\shownote{Accessed: [2025.9.9]. [Online]}.


\bibitem[Liu et~al\mbox{.}(2023)]%
        {liu2023rise}
\bibfield{author}{\bibinfo{person}{Xuanzhe Liu}, \bibinfo{person}{Diandian Gu}, \bibinfo{person}{Zhenpeng Chen}, \bibinfo{person}{Jinfeng Wen}, \bibinfo{person}{Zili Zhang}, \bibinfo{person}{Yun Ma}, \bibinfo{person}{Haoyu Wang}, {and} \bibinfo{person}{Xin Jin}.} \bibinfo{year}{2023}\natexlab{}.
\newblock \showarticletitle{Rise of distributed deep learning training in the big model era: From a software engineering perspective}.
\newblock \bibinfo{journal}{\emph{ACM Transactions on Software Engineering and Methodology}} \bibinfo{volume}{32}, \bibinfo{number}{6} (\bibinfo{year}{2023}), \bibinfo{pages}{1--26}.
\newblock


\bibitem[Long and Chen(2022)]%
        {long2022reporting}
\bibfield{author}{\bibinfo{person}{Guoming Long} {and} \bibinfo{person}{Tao Chen}.} \bibinfo{year}{2022}\natexlab{}.
\newblock \showarticletitle{On reporting performance and accuracy bugs for deep learning frameworks: An exploratory study from github}. In \bibinfo{booktitle}{\emph{Proceedings of the 26th international conference on evaluation and assessment in software engineering}}. \bibinfo{pages}{90--99}.
\newblock


\bibitem[MantasLukauskas(2021)]%
        {deepspeed_issue1342}
\bibfield{author}{\bibinfo{person}{MantasLukauskas}.} \bibinfo{year}{2021}\natexlab{}.
\newblock \bibinfo{title}{Issue 1342 in DeepSpeed}.
\newblock \bibinfo{howpublished}{Available: \url{https://github.com/deepspeedai/DeepSpeed/issues/1342}}.
\newblock
\newblock
\shownote{Accessed: [2025.9.9]. [Online]}.


\bibitem[McHugh(2012)]%
        {mchugh2012interrater}
\bibfield{author}{\bibinfo{person}{Mary~L McHugh}.} \bibinfo{year}{2012}\natexlab{}.
\newblock \showarticletitle{Interrater reliability: the kappa statistic}.
\newblock \bibinfo{journal}{\emph{Biochemia medica}} \bibinfo{volume}{22}, \bibinfo{number}{3} (\bibinfo{year}{2012}), \bibinfo{pages}{276--282}.
\newblock


\bibitem[MikeChenfu(2023)]%
        {deepspeed_issue2672}
\bibfield{author}{\bibinfo{person}{MikeChenfu}.} \bibinfo{year}{2023}\natexlab{}.
\newblock \bibinfo{title}{Issue 2672 in DeepSpeed}.
\newblock \bibinfo{howpublished}{Available: \url{https://github.com/deepspeedai/DeepSpeed/issues/2672}}.
\newblock
\newblock
\shownote{Accessed: [2025.9.9]. [Online]}.


\bibitem[Modas-Li(2023)]%
        {deepspeed_issue3340}
\bibfield{author}{\bibinfo{person}{Modas-Li}.} \bibinfo{year}{2023}\natexlab{}.
\newblock \bibinfo{title}{Issue 3340 in DeepSpeed}.
\newblock \bibinfo{howpublished}{Available: \url{https://github.com/deepspeedai/DeepSpeed/issues/3340}}.
\newblock
\newblock
\shownote{Accessed: [2025.9.9]. [Online]}.


\bibitem[nameless0704(2023)]%
        {colossalAI_issue2941}
\bibfield{author}{\bibinfo{person}{nameless0704}.} \bibinfo{year}{2023}\natexlab{}.
\newblock \bibinfo{title}{Issue 2941 in colossalAI}.
\newblock \bibinfo{howpublished}{Available: \url{https://github.com/hpcaitech/ColossalAI/issues/2941}}.
\newblock
\newblock
\shownote{Accessed: [2025.9.9]. [Online]}.


\bibitem[Narayanan et~al\mbox{.}(2021)]%
        {narayanan2021efficient}
\bibfield{author}{\bibinfo{person}{Deepak Narayanan}, \bibinfo{person}{Mohammad Shoeybi}, \bibinfo{person}{Jared Casper}, \bibinfo{person}{Patrick LeGresley}, \bibinfo{person}{Mostofa Patwary}, \bibinfo{person}{Vijay Korthikanti}, \bibinfo{person}{Dmitri Vainbrand}, \bibinfo{person}{Prethvi Kashinkunti}, \bibinfo{person}{Julie Bernauer}, \bibinfo{person}{Bryan Catanzaro}, {et~al\mbox{.}}} \bibinfo{year}{2021}\natexlab{}.
\newblock \showarticletitle{Efficient large-scale language model training on gpu clusters using megatron-lm}. In \bibinfo{booktitle}{\emph{Proceedings of the international conference for high performance computing, networking, storage and analysis}}. \bibinfo{pages}{1--15}.
\newblock


\bibitem[Park et~al\mbox{.}(2020)]%
        {park2020hetpipe}
\bibfield{author}{\bibinfo{person}{Jay~H Park}, \bibinfo{person}{Gyeongchan Yun}, \bibinfo{person}{M~Yi Chang}, \bibinfo{person}{Nguyen~T Nguyen}, \bibinfo{person}{Seungmin Lee}, \bibinfo{person}{Jaesik Choi}, \bibinfo{person}{Sam~H Noh}, {and} \bibinfo{person}{Young-ri Choi}.} \bibinfo{year}{2020}\natexlab{}.
\newblock \showarticletitle{$\{$HetPipe$\}$: Enabling large $\{$DNN$\}$ training on (whimpy) heterogeneous $\{$GPU$\}$ clusters through integration of pipelined model parallelism and data parallelism}. In \bibinfo{booktitle}{\emph{2020 USENIX Annual Technical Conference (USENIX ATC 20)}}. \bibinfo{pages}{307--321}.
\newblock


\bibitem[Paszke et~al\mbox{.}(2019)]%
        {paszke2019pytorch}
\bibfield{author}{\bibinfo{person}{Adam Paszke}, \bibinfo{person}{Sam Gross}, \bibinfo{person}{Francisco Massa}, \bibinfo{person}{Adam Lerer}, \bibinfo{person}{James Bradbury}, \bibinfo{person}{Gregory Chanan}, \bibinfo{person}{Trevor Killeen}, \bibinfo{person}{Zeming Lin}, \bibinfo{person}{Natalia Gimelshein}, \bibinfo{person}{Luca Antiga}, {et~al\mbox{.}}} \bibinfo{year}{2019}\natexlab{}.
\newblock \showarticletitle{Pytorch: An imperative style, high-performance deep learning library}.
\newblock \bibinfo{journal}{\emph{Advances in neural information processing systems}}  \bibinfo{volume}{32} (\bibinfo{year}{2019}).
\newblock


\bibitem[philschmid(2022a)]%
        {deepspeed_issue1766}
\bibfield{author}{\bibinfo{person}{philschmid}.} \bibinfo{year}{2022}\natexlab{a}.
\newblock \bibinfo{title}{Issue 1766 in DeepSpeed}.
\newblock \bibinfo{howpublished}{Available: \url{https://github.com/deepspeedai/DeepSpeed/issues/1766}}.
\newblock
\newblock
\shownote{Accessed: [2025.9.9]. [Online]}.


\bibitem[philschmid(2022b)]%
        {deepspeed_issue1772}
\bibfield{author}{\bibinfo{person}{philschmid}.} \bibinfo{year}{2022}\natexlab{b}.
\newblock \bibinfo{title}{Issue 1772 in DeepSpeed}.
\newblock \bibinfo{howpublished}{Available: \url{https://github.com/deepspeedai/DeepSpeed/issues/1772}}.
\newblock
\newblock
\shownote{Accessed: [2025.9.9]. [Online]}.


\bibitem[powermano(2022a)]%
        {colossalAI_issue1035}
\bibfield{author}{\bibinfo{person}{powermano}.} \bibinfo{year}{2022}\natexlab{a}.
\newblock \bibinfo{title}{Issue 1035 in colossalAI}.
\newblock \bibinfo{howpublished}{Available: \url{https://github.com/hpcaitech/ColossalAI/issues/1035}}.
\newblock
\newblock
\shownote{Accessed: [2025.9.9]. [Online]}.


\bibitem[powermano(2022b)]%
        {colossalAI_issue1082}
\bibfield{author}{\bibinfo{person}{powermano}.} \bibinfo{year}{2022}\natexlab{b}.
\newblock \bibinfo{title}{Issue 1082 in colossalAI}.
\newblock \bibinfo{howpublished}{Available: \url{https://github.com/hpcaitech/ColossalAI/issues/1082}}.
\newblock
\newblock
\shownote{Accessed: [2025.9.9]. [Online]}.


\bibitem[Rajbhandari et~al\mbox{.}(2020)]%
        {rajbhandari2020zero}
\bibfield{author}{\bibinfo{person}{Samyam Rajbhandari}, \bibinfo{person}{Jeff Rasley}, \bibinfo{person}{Olatunji Ruwase}, {and} \bibinfo{person}{Yuxiong He}.} \bibinfo{year}{2020}\natexlab{}.
\newblock \showarticletitle{Zero: Memory optimizations toward training trillion parameter models}. In \bibinfo{booktitle}{\emph{SC20: International Conference for High Performance Computing, Networking, Storage and Analysis}}. IEEE, \bibinfo{pages}{1--16}.
\newblock


\bibitem[Rasley et~al\mbox{.}(2020)]%
        {rasley2020deepspeed}
\bibfield{author}{\bibinfo{person}{Jeff Rasley}, \bibinfo{person}{Samyam Rajbhandari}, \bibinfo{person}{Olatunji Ruwase}, {and} \bibinfo{person}{Yuxiong He}.} \bibinfo{year}{2020}\natexlab{}.
\newblock \showarticletitle{Deepspeed: System optimizations enable training deep learning models with over 100 billion parameters}. In \bibinfo{booktitle}{\emph{Proceedings of the 26th ACM SIGKDD international conference on knowledge discovery \& data mining}}. \bibinfo{pages}{3505--3506}.
\newblock


\bibitem[Riraifu(2023)]%
        {deepspeed_issue4829}
\bibfield{author}{\bibinfo{person}{Riraifu}.} \bibinfo{year}{2023}\natexlab{}.
\newblock \bibinfo{title}{Issue 4829 in DeepSpeed}.
\newblock \bibinfo{howpublished}{Available: \url{https://github.com/deepspeedai/DeepSpeed/issues/4829}}.
\newblock
\newblock
\shownote{Accessed: [2025.9.9]. [Online]}.


\bibitem[Sanger2000(2022)]%
        {deepspeed_issue1923}
\bibfield{author}{\bibinfo{person}{Sanger2000}.} \bibinfo{year}{2022}\natexlab{}.
\newblock \bibinfo{title}{Issue 1923 in DeepSpeed}.
\newblock \bibinfo{howpublished}{Available: \url{https://github.com/deepspeedai/DeepSpeed/issues/1923}}.
\newblock
\newblock
\shownote{Accessed: [2025.9.9]. [Online]}.


\bibitem[Sergeev and Del~Balso(2018)]%
        {sergeev2018horovod}
\bibfield{author}{\bibinfo{person}{Alexander Sergeev} {and} \bibinfo{person}{Mike Del~Balso}.} \bibinfo{year}{2018}\natexlab{}.
\newblock \showarticletitle{Horovod: fast and easy distributed deep learning in TensorFlow}.
\newblock \bibinfo{journal}{\emph{arXiv preprint arXiv:1802.05799}} (\bibinfo{year}{2018}).
\newblock


\bibitem[Shah et~al\mbox{.}(2025)]%
        {shah2025towards}
\bibfield{author}{\bibinfo{person}{Mehil~B Shah}, \bibinfo{person}{Mohammad~Masudur Rahman}, {and} \bibinfo{person}{Foutse Khomh}.} \bibinfo{year}{2025}\natexlab{}.
\newblock \showarticletitle{Towards enhancing the reproducibility of deep learning bugs: an empirical study}.
\newblock \bibinfo{journal}{\emph{Empirical Software Engineering}} \bibinfo{volume}{30}, \bibinfo{number}{1} (\bibinfo{year}{2025}), \bibinfo{pages}{23}.
\newblock


\bibitem[Shanahan(2024)]%
        {shanahan2024talking}
\bibfield{author}{\bibinfo{person}{Murray Shanahan}.} \bibinfo{year}{2024}\natexlab{}.
\newblock \showarticletitle{Talking about large language models}.
\newblock \bibinfo{journal}{\emph{Commun. ACM}} \bibinfo{volume}{67}, \bibinfo{number}{2} (\bibinfo{year}{2024}), \bibinfo{pages}{68--79}.
\newblock


\bibitem[Shoeybi et~al\mbox{.}(2019)]%
        {shoeybi2019megatron}
\bibfield{author}{\bibinfo{person}{Mohammad Shoeybi}, \bibinfo{person}{Mostofa Patwary}, \bibinfo{person}{Raul Puri}, \bibinfo{person}{Patrick LeGresley}, \bibinfo{person}{Jared Casper}, {and} \bibinfo{person}{Bryan Catanzaro}.} \bibinfo{year}{2019}\natexlab{}.
\newblock \showarticletitle{Megatron-lm: Training multi-billion parameter language models using model parallelism}.
\newblock \bibinfo{journal}{\emph{arXiv preprint arXiv:1909.08053}} (\bibinfo{year}{2019}).
\newblock


\bibitem[SimZhou(2022)]%
        {colossalAI_issue990}
\bibfield{author}{\bibinfo{person}{SimZhou}.} \bibinfo{year}{2022}\natexlab{}.
\newblock \bibinfo{title}{Issue 990 in colossalAI}.
\newblock \bibinfo{howpublished}{Available: \url{https://github.com/hpcaitech/ColossalAI/issues/990}}.
\newblock
\newblock
\shownote{Accessed: [2025.9.9]. [Online]}.


\bibitem[snowyday(2024)]%
        {deepspeed_issue5278}
\bibfield{author}{\bibinfo{person}{snowyday}.} \bibinfo{year}{2024}\natexlab{}.
\newblock \bibinfo{title}{Issue 5278 in DeepSpeed}.
\newblock \bibinfo{howpublished}{Available: \url{https://github.com/deepspeedai/DeepSpeed/issues/5278}}.
\newblock
\newblock
\shownote{Accessed: [2025.9.9]. [Online]}.


\bibitem[songdezhao(2025)]%
        {deepspeed_issue7278}
\bibfield{author}{\bibinfo{person}{songdezhao}.} \bibinfo{year}{2025}\natexlab{}.
\newblock \bibinfo{title}{Issue 7278 in DeepSpeed}.
\newblock \bibinfo{howpublished}{Available: \url{https://github.com/deepspeedai/DeepSpeed/issues/7278}}.
\newblock
\newblock
\shownote{Accessed: [2025.9.9]. [Online]}.


\bibitem[sunxiaoyu12(2023)]%
        {deepspeed_issue4034}
\bibfield{author}{\bibinfo{person}{sunxiaoyu12}.} \bibinfo{year}{2023}\natexlab{}.
\newblock \bibinfo{title}{Issue 4034 in DeepSpeed}.
\newblock \bibinfo{howpublished}{Available: \url{https://github.com/deepspeedai/DeepSpeed/issues/4034}}.
\newblock
\newblock
\shownote{Accessed: [2025.9.9]. [Online]}.


\bibitem[syngokhan(2023)]%
        {deepspeed_issue4648}
\bibfield{author}{\bibinfo{person}{syngokhan}.} \bibinfo{year}{2023}\natexlab{}.
\newblock \bibinfo{title}{Issue 4648 in DeepSpeed}.
\newblock \bibinfo{howpublished}{Available: \url{https://github.com/deepspeedai/DeepSpeed/issues/4648}}.
\newblock
\newblock
\shownote{Accessed: [2025.9.9]. [Online]}.


\bibitem[Tambon et~al\mbox{.}(2024)]%
        {tambon2024silent}
\bibfield{author}{\bibinfo{person}{Florian Tambon}, \bibinfo{person}{Amin Nikanjam}, \bibinfo{person}{Le An}, \bibinfo{person}{Foutse Khomh}, {and} \bibinfo{person}{Giuliano Antoniol}.} \bibinfo{year}{2024}\natexlab{}.
\newblock \showarticletitle{Silent bugs in deep learning frameworks: an empirical study of keras and tensorflow}.
\newblock \bibinfo{journal}{\emph{Empirical Software Engineering}} \bibinfo{volume}{29}, \bibinfo{number}{1} (\bibinfo{year}{2024}), \bibinfo{pages}{10}.
\newblock


\bibitem[theblackcat102(2022)]%
        {colossalAI_issue2145}
\bibfield{author}{\bibinfo{person}{theblackcat102}.} \bibinfo{year}{2022}\natexlab{}.
\newblock \bibinfo{title}{Issue 2145 in colossalAI}.
\newblock \bibinfo{howpublished}{Available: \url{https://github.com/hpcaitech/ColossalAI/issues/2145}}.
\newblock
\newblock
\shownote{Accessed: [2025.9.9]. [Online]}.


\bibitem[tohneecao(2023)]%
        {deepspeed_issue3360}
\bibfield{author}{\bibinfo{person}{tohneecao}.} \bibinfo{year}{2023}\natexlab{}.
\newblock \bibinfo{title}{Issue 3360 in DeepSpeed}.
\newblock \bibinfo{howpublished}{Available: \url{https://github.com/deepspeedai/DeepSpeed/issues/3360}}.
\newblock
\newblock
\shownote{Accessed: [2025.9.9]. [Online]}.


\bibitem[uygnef(2023)]%
        {deepspeed_issue3472}
\bibfield{author}{\bibinfo{person}{uygnef}.} \bibinfo{year}{2023}\natexlab{}.
\newblock \bibinfo{title}{Issue 3472 in DeepSpeed}.
\newblock \bibinfo{howpublished}{Available: \url{https://github.com/deepspeedai/DeepSpeed/issues/3472}}.
\newblock
\newblock
\shownote{Accessed: [2025.9.9]. [Online]}.


\bibitem[Viera et~al\mbox{.}(2005)]%
        {viera2005understanding}
\bibfield{author}{\bibinfo{person}{Anthony~J Viera}, \bibinfo{person}{Joanne~M Garrett}, {et~al\mbox{.}}} \bibinfo{year}{2005}\natexlab{}.
\newblock \showarticletitle{Understanding interobserver agreement: the kappa statistic}.
\newblock \bibinfo{journal}{\emph{Fam med}} \bibinfo{volume}{37}, \bibinfo{number}{5} (\bibinfo{year}{2005}), \bibinfo{pages}{360--363}.
\newblock


\bibitem[Vijayaraghavan and Kaner(2003)]%
        {vijayaraghavan2003bug}
\bibfield{author}{\bibinfo{person}{Giri Vijayaraghavan} {and} \bibinfo{person}{Cem Kaner}.} \bibinfo{year}{2003}\natexlab{}.
\newblock \showarticletitle{Bug taxonomies: Use them to generate better tests}.
\newblock \bibinfo{journal}{\emph{Star East}}  \bibinfo{volume}{2003} (\bibinfo{year}{2003}), \bibinfo{pages}{1--40}.
\newblock


\bibitem[Wang et~al\mbox{.}(2023)]%
        {wang2023compatibility}
\bibfield{author}{\bibinfo{person}{Jun Wang}, \bibinfo{person}{Guanping Xiao}, \bibinfo{person}{Shuai Zhang}, \bibinfo{person}{Huashan Lei}, \bibinfo{person}{Yepang Liu}, {and} \bibinfo{person}{Yulei Sui}.} \bibinfo{year}{2023}\natexlab{}.
\newblock \showarticletitle{Compatibility issues in deep learning systems: Problems and opportunities}. In \bibinfo{booktitle}{\emph{Proceedings of the 31st ACM Joint European Software Engineering Conference and Symposium on the Foundations of Software Engineering}}. \bibinfo{pages}{476--488}.
\newblock


\bibitem[willfant(2022)]%
        {colossalAI_issue1051}
\bibfield{author}{\bibinfo{person}{willfant}.} \bibinfo{year}{2022}\natexlab{}.
\newblock \bibinfo{title}{Issue 1051 in colossalAI}.
\newblock \bibinfo{howpublished}{Available: \url{https://github.com/hpcaitech/ColossalAI/issues/1051}}.
\newblock
\newblock
\shownote{Accessed: [2025.9.9]. [Online]}.


\bibitem[winktool(2023)]%
        {colossalAI_issue2605}
\bibfield{author}{\bibinfo{person}{winktool}.} \bibinfo{year}{2023}\natexlab{}.
\newblock \bibinfo{title}{Issue 2605 in colossalAI}.
\newblock \bibinfo{howpublished}{Available: \url{https://github.com/hpcaitech/ColossalAI/issues/2605}}.
\newblock
\newblock
\shownote{Accessed: [2025.9.9]. [Online]}.


\bibitem[withyou1771(2023)]%
        {colossalAI_issue3202}
\bibfield{author}{\bibinfo{person}{withyou1771}.} \bibinfo{year}{2023}\natexlab{}.
\newblock \bibinfo{title}{Issue 3202 in colossalAI}.
\newblock \bibinfo{howpublished}{Available: \url{https://github.com/hpcaitech/ColossalAI/issues/3202}}.
\newblock
\newblock
\shownote{Accessed: [2025.9.9]. [Online]}.


\bibitem[Yang et~al\mbox{.}(2025)]%
        {yang2025towards}
\bibfield{author}{\bibinfo{person}{Haowen Yang}, \bibinfo{person}{Zhengda Li}, \bibinfo{person}{Zhiqing Zhong}, \bibinfo{person}{Xiaoying Tang}, {and} \bibinfo{person}{Pinjia He}.} \bibinfo{year}{2025}\natexlab{}.
\newblock \showarticletitle{Towards Understanding Performance Bugs in Popular Data Science Libraries}.
\newblock \bibinfo{journal}{\emph{Proceedings of the ACM on Software Engineering}} \bibinfo{volume}{2}, \bibinfo{number}{FSE} (\bibinfo{year}{2025}), \bibinfo{pages}{2335--2358}.
\newblock


\bibitem[Yang et~al\mbox{.}(2022a)]%
        {yang2022comprehensive}
\bibfield{author}{\bibinfo{person}{Yilin Yang}, \bibinfo{person}{Tianxing He}, \bibinfo{person}{Zhilong Xia}, {and} \bibinfo{person}{Yang Feng}.} \bibinfo{year}{2022}\natexlab{a}.
\newblock \showarticletitle{A comprehensive empirical study on bug characteristics of deep learning frameworks}.
\newblock \bibinfo{journal}{\emph{Information and Software Technology}}  \bibinfo{volume}{151} (\bibinfo{year}{2022}), \bibinfo{pages}{107004}.
\newblock


\bibitem[Yang et~al\mbox{.}(2022b)]%
        {yang2022survey}
\bibfield{author}{\bibinfo{person}{Yanming Yang}, \bibinfo{person}{Xin Xia}, \bibinfo{person}{David Lo}, {and} \bibinfo{person}{John Grundy}.} \bibinfo{year}{2022}\natexlab{b}.
\newblock \showarticletitle{A survey on deep learning for software engineering}.
\newblock \bibinfo{journal}{\emph{ACM Computing Surveys (CSUR)}} \bibinfo{volume}{54}, \bibinfo{number}{10s} (\bibinfo{year}{2022}), \bibinfo{pages}{1--73}.
\newblock


\bibitem[Zhang et~al\mbox{.}(2020)]%
        {zhang2020empirical}
\bibfield{author}{\bibinfo{person}{Ru Zhang}, \bibinfo{person}{Wencong Xiao}, \bibinfo{person}{Hongyu Zhang}, \bibinfo{person}{Yu Liu}, \bibinfo{person}{Haoxiang Lin}, {and} \bibinfo{person}{Mao Yang}.} \bibinfo{year}{2020}\natexlab{}.
\newblock \showarticletitle{An empirical study on program failures of deep learning jobs}. In \bibinfo{booktitle}{\emph{Proceedings of the ACM/IEEE 42nd international conference on software engineering}}. \bibinfo{pages}{1159--1170}.
\newblock


\bibitem[Zhang et~al\mbox{.}(2018)]%
        {zhang2018empirical}
\bibfield{author}{\bibinfo{person}{Yuhao Zhang}, \bibinfo{person}{Yifan Chen}, \bibinfo{person}{Shing-Chi Cheung}, \bibinfo{person}{Yingfei Xiong}, {and} \bibinfo{person}{Lu Zhang}.} \bibinfo{year}{2018}\natexlab{}.
\newblock \showarticletitle{An empirical study on tensorflow program bugs}. In \bibinfo{booktitle}{\emph{Proceedings of the 27th ACM SIGSOFT international symposium on software testing and analysis}}. \bibinfo{pages}{129--140}.
\newblock


\end{thebibliography}

\end{document}